\documentclass[10pt]{article}
\usepackage[totalwidth=450pt, totalheight=590pt, centering]{geometry}

\usepackage{amsmath,amssymb,mathrsfs}
\usepackage{bbm} 

\usepackage[dvips]{graphicx}

\usepackage{pstricks}

\usepackage[dvips,bookmarks=true]{hyperref} 
\hypersetup{pdfstartview=FitH,pdfhighlight=/O,colorlinks=false}

\newcommand{\be}{\begin{equation}}
\newcommand{\ee}{\end{equation}}
\newcommand{\ben}{\begin{equation*}}
\newcommand{\een}{\end{equation*}}

\newcommand{\mc}[1]{\mathcal{#1}}

\newcommand{\de}{\delta}
\newcommand{\p}{\partial}

\newcommand{\eps}{\varepsilon}

\newcommand{\ph}[1]{\phantom{#1}}
\newcommand{\mdots}{,.\,.\,,}

\title{\Large{\textbf{The length operator in Loop Quantum Gravity}}}

\author{Eugenio Bianchi\footnote{\texttt{e.bianchi@sns.it}}\\[.35em]
\small{Scuola  Normale Superiore,
Piazza dei Cavalieri 7, I-56126 Pisa,  Italy }\\
\small{Centre de Physique Th\'eorique de Luminy, Universit\'e de la M\'editerran\'ee, F-13288 Marseille, France}
}

\date{\small \today}

\begin{document}

\maketitle

\begin{abstract}
The dual picture of quantum geometry provided by a spin~network state is discussed. From this perspective, we introduce a new operator in Loop Quantum Gravity -- the length operator. 
We describe its quantum geometrical meaning and derive some of its properties.
In particular we show that the operator has a discrete spectrum and is diagonalized by appropriate superpositions 
 of spin~network states. A series of eigenstates and eigenvalues is presented and an explicit check of its semiclassical properties is discussed.

\begin{flushleft}
\end{flushleft}
\begin{flushleft}
\end{flushleft}
\end{abstract}


\vspace{3em}

A remarkable feature of the loop approach \cite{Rovelli:2004tv,Thiemann:book2007,Ashtekar:2004eh} to the problem of quantum gravity \cite{Oriti:book2009} is the prediction of a quantum discreteness of space at the Planck scale. Such discreteness manifests itself in the analysis of the spectrum of geometric operators describing the volume of a region of space \cite{Rovelli:1994ge, Ashtekar:1997fb} or the area of a surface separating two such regions \cite{Rovelli:1994ge,Ashtekar:1996eg}. In this paper we introduce a new operator -- the length operator -- study its properties 
and show that it has a discrete spectrum and an appropriate semiclassical behaviour.  For a different attempt to introduce a length operator in Loop Quantum Gravity, see Thiemann's paper \cite{Thiemann:1996at}. For some remarks about why it is difficult to introduce a length operator is Loop Quantum Gravity see the review \cite{Rovelli:1997yv}. In the following we describe the picture of quantum geometry coming from Loop Quantum Gravity and the role played by the length in this picture (section \ref{sec:picture}), we recall the standard procedure used in Loop Quantum Gravity when introducing an operator corresponding to a given \emph{classical}  geometrical quantity (section \ref{sec:quantization}), we point out the difficulties to overcome in order to introduce the length operator (section \ref{sec:difficulties}), discuss the strategy that we follow (sections \ref{sec:strategy}-\ref{sec:construction subsec}) and finally present our results in sections \ref{sec:properties} and \ref{sec:extended length}.

\section{The dual picture of quantum geometry}\label{sec:picture}
\begin{figure}
\centering
\includegraphics[width=0.25\textwidth]{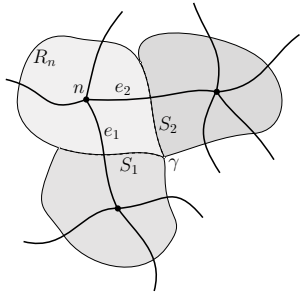}
\caption{A portion of a spin~network graph and the associated dual picture of quantum  geometry. The region $R_n$ is dual to the node $n$. Two adjacent regions are shown. The surfaces $S_1$ and $S_2$ are dual to the links $e_1$ and $e_2$. They identify a curve $\gamma$ on the boundary of $R_n$.}
\label{fig:dual picture}
\end{figure}

In Loop Quantum Gravity, the state of the 3-geometry can be given in terms of a linear superposition of spin~network states. Such spin~network states consist of a graph embedded in a 3-manifold and a coloring of its edges and its nodes in terms of SU(2) irreducible representations and of SU(2) intertwiners. Thanks to the existence of a volume operator and an area operator, the following \emph{dual} picture of the quantum geometry of a spin~network state is available (see sections 1.2.2 and 6.7 of \cite{Rovelli:2004tv} for a detailed discussion): a node of the spin~network corresponds to a chunk of space with definite volume while a link connecting two nodes corresponds to an interface of definite area which separates two chunks (see figure \ref{fig:dual picture}). Moreover, \emph{a node connected to two other nodes identifies two surfaces which intersect at a curve}. The operator we introduce in this paper corresponds to the length of this curve.

\section{Quantization of 3-geometric observables}\label{sec:quantization}
At the classical level -- in general relativity -- volume and area are functions of the metric which is the dynamical variable. Generally speaking, in ``quantum geometry'' approaches \cite{Isham:1992ms}, the metric is promoted to an operator on a Hilbert space. Therefore, admitting to have enough mathematical control on the theory, one can introduce for instance a volume operator as a function of the metric operator and study its eigenstates and its spectrum. Such mathematical control is available in Loop Quantum Gravity thanks to the existence of the Ashtekar-Lewandowski measure on the space of connections \cite{Ashtekar:1993wf,Ashtekar:1994wa,Lewandowski:2005jk,Fleischhack:2004jc}.

The starting point for quantization is canonical general relativity written in terms of a real $SU(2)$ connection\footnote{The following notational conventions are adopted: $a,b,\ldots$ are \emph{space} indices while $i,j,\ldots$ are \emph{internal} $SU(2)$ indices so that we have $A(x)=A_a^i(x)\, T_i\,dx^a\,$ with $T_i$ a generator of $SU(2)$.} $A_a^i(x)$ and its conjugate momentum $E^a_i(x)$, i.e. in terms of the so-called Ashtekar-Barbero variables with real Immirzi parameter $\gamma$ \cite{Ashtekar:1986yd,Ashtekar:1987gu,Barbero:1994ap,Immirzi:1996di}. All the information about the geometry of space is encoded in the field $E^a_i(x)$. Being the momentum conjugate to a connection, it deserves the name \emph{electric field}. It is a density of weight one which corresponds to the inverse densitized triad, $E^a_i=\frac{1}{2}\eps_{ijk}\eps^{abc} e^j_b e^k_c\,$. 
 For instance, the volume of a region $R$ is a functional of the electric field $E^a_i(x)$ and is given by
\begin{equation}
V(R)=\int_R \!d^3	x\;\sqrt{\frac{1}{3!}|\eps^{ijk}\eps_{abc}E^a_i E^b_j E^c_k|}\;.
	\label{eq:vol}
\end{equation}

Now we move on to the quantum theory \cite{Rovelli:1987df,Rovelli:1989za}. The essential assumption in Loop Quantum Gravity is that the mathematically well-defined operators acting on the Hilbert space are the holonomy\footnote{We recall that, at the classical level, the holonomy along a curve $e$ is given in terms of the connection $A_a^i(x)$ by the following path ordered exponential
\begin{equation*}
h_e[A]=\mc{P}\exp\, i\! \int_e	A_a^i(x)\, T_i\, dx^a\;.
\end{equation*}} of the connection along a curve $e$, $h_e[A]$, and the smearing of the two-form $E_i=E^a_i(x) \;\eps_{abc}dx^b\wedge dx^c$ over a surface $S$, namely the flux of the electric field through the surface,
\begin{equation}
	F_i(S)=\int_S E^a_i(x) \;\eps_{abc}dx^b\wedge dx^c\;.
	\label{eq:flux}
\end{equation}
Every operator is to be considered as a function of such fundamental quantities\footnote{This crucial assumption of having well-defined operators resulting from the smearing of $n$-forms over $n$-manifolds leads to a remarkable interplay between functional analysis and differential geometry. Notice however that this is not something due: for instance in quantum electrodynamics both the connection and the electric field are taken with smearing over $3$-dimensional regions. 
 The motivation behind this assumption in Loop Quantum Gravity comes from the covariance properties of the two quantities $h_e[A]$ and $F_i(S)$ under diffeomorphisms of the $3$-manifold $\Sigma$. Such mathematical assumption has far reaching consequence in terms of physical predictions of the theory.}.

Let's recall in a few lines some basics. The Hilbert space of states of the theory ($\mc{K}$ in the nomenclature of \cite{Rovelli:2004tv}) is the space of functionals of the connections $\Psi[A]$ which are square integrable with respect to the Ashtekar-Lewandowski measure. Cylindrical functions, i.e. functions which depend on the connection only through its holonomy along the edges of a graph embedded in $\Sigma$, are dense in the Hilbert space. On cylindrical functions the Ashtekar-Lewandowski measure has the property of reducing to a product of Haar measures on $SU(2)$. In Loop Quantum Gravity we are interested in the space $\mc{K}_0$ of $SU(2)$-gauge-invariant functionals of the connection. At the level of cylindrical functions this amounts to restricting attention to the class of functions invariant under $SU(2)$ transformations at the nodes of the graph. Spin~network states \cite{Rovelli:1995ac,Baez:1994hx} provide an orthonormal basis of $\mc{K}_0$. They are defined in the following way. We consider a closed graph $\Gamma$ embedded in the $3$-manifold $\Sigma$. To each edges $e$ of the graph we associate an irreducible representation $j_e$ of $SU(2)$. To each nodes $n$ of the graph we associate an invariant vector, also called an intertwiner, $i_n$ in the tensor product of representations labelling the edges converging at the node. Then the spin~network state labelled by the triple $\{\Gamma,j_e,i_n\}$ is given by the following product over nodes and links of $\Gamma$
\begin{equation}
	\Psi_{\Gamma,j,i}[A]=\bigotimes_{n\subset\Gamma} v_{i_n} \bigotimes_{e\subset\Gamma}\mc{D}^{(j_e)}(h_e[A])
	\label{eq:spin network state}
\end{equation}
where $\mc{D}^{(j)}(h_e[A])$ is the representation $j$ of the holonomy of the connection along the edge $e$. Spin~network states with graph $\Gamma$ span a subspace of $\mc{K}_0$. We call this subspace $\mc{K}_0(\Gamma)$.
 
 Let's introduce some graphical notation. We represent (a) the holonomy $h_e[A]$ in representation $j$ along a curve $e$ embedded in the $3$-manifold $\Sigma$ by a labelled edge
and (b) an intertwining tensor $v_{i_n}$, 
that is the tensor associated to an invariant vector $i_n$ in the tensor product of $L$ irreducible $SU(2)$ representations, by a $L$-valent node 

\vspace{-1.5em}

\begin{equation}
\mc{D}^{(j)}(h_e[A])_m^{\ph{m}m'}=\;\;\parbox[c]{100pt}{\includegraphics{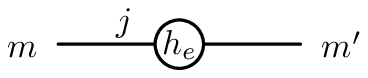}}  \hspace{4em} 
	{v_{i_n}}^{(j_1\ldots j_L)}_{m_1\ldots m_L}=\;\;\parbox[c]{80pt}{\includegraphics{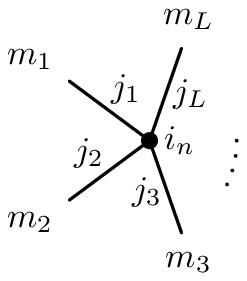}}\;\;.
	\label{eq:holonomy and intertwiner diagram}
\end{equation}
A spin~network state can be represented diagrammatically using the building blocks (a) and (b).

The action of the fundamental operators $\widehat{h_e}$ and $\widehat{F_i}(S)$ is easy to declare on cylindrical functions and can be defined on the whole Hilbert space through self-adjoint extension. The holonomy operator acts multiplicatively on cylindrical functions. Thanks to the fact that a cylindrical function can be expanded on the spin~network basis (Peter-Weyl theorem), the action of the flux operator on a cylindrical function is completely determined by its action on (the representation $j$ of) the holonomy along a curve $e$. For a surface $S$ which intersects a curve $e$ splitting it into two segments $b$ and $c$, we have  
\begin{equation}
	\widehat{F_i}(S)\;\mc{D}^{(j)}(h_e[A])_m^{\ph{m}n}=\pm\,8\pi \gamma L_P^2 \;\mc{D}^{(j)}(h_b[A])_m^{\ph{m}m'}\;T^{(j)\;n'}_{i\,m'}\;\mc{D}^{(j)}(h_c[A])_{n'}^{\ph{n\,}n}
	\label{eq:grasping formula}
\end{equation}
where $T^{(j)\;n}_{i\,m}$ is a traceless hermitian matrix given by the representation $j$ of the $SU(2)$ generator $T_i$, $\gamma$ is the Immirzi parameter and $L_P$ is Planck length\footnote{Expression (\ref{eq:grasping formula}) corresponds to requiring the commutator of the holonomy operator with the flux operator to be equal to $i \hbar$ times the Poisson-bracket of the holonomy with the flux computed at the classical level. That is quantizing canonically the classical expression $\{F_i(S),h_e[A]\}$. The $8\pi \gamma L_P^2$ in (\ref{eq:grasping formula}) comes from the basic Poisson-bracket
\begin{equation}
\{A_a^i(x),E^b_j(x')\}=	\gamma \frac{8\pi G_N}{c^3}\, \delta^i_{\ph{i}j}\,\delta_a^{\ph{a}b}\,\delta(x-x')\;.
\label{eq:poisson bracket A,E}
\end{equation}
}. The sign is dictated by the relative orientation of the surface $S$ and the edge $e$. In case there is no intersection or if the edge lies in the surface, then the flux operator annihilates the state.

The action of the flux operator for a surface dual to a link of a spin~network state is generally called a grasping. Equation (\ref{eq:grasping formula}) plays a key role in the construction of geometric operators. It can be represented diagrammatically in the following way:

\vspace{-1em}

\begin{equation}
\widehat{F_i}(S)\;\;\parbox[c]{100pt}{\includegraphics{feyn.holonomy.eps}}\;\; =\;\pm\,8\pi \gamma L_P^2 \;\;\parbox[c]{160pt}{\includegraphics{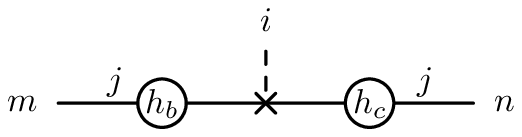}\\[5pt]}\;\;.
	\label{eq:grasping diagram}
\end{equation}

\vspace{-1em}

\noindent where we have represented with a dashed line a link in the adjoint representation $j=1$.

\subsection{The volume operator}

Here we describe in some detail the construction of the volume operator as we will follow similar steps when introducing the length operator in next section. In order to introduce in the quantum theory an operator corresponding to the volume of a region, the starting point is the classical expression (\ref{eq:vol}). Applying the canonical quantization procedure, however, is not straightforward: at the quantum level the well-defined operator representing the geometry of space is not $E^a_i(x)$ but its flux through a surface. Therefore the quantization strategy is to find a regularized expression for the classical volume in terms of fluxes, to promote this expression to an operator and then analyze the existence of the limit in the Hilbert space topology. If the limit exists, then we can say that we have a \emph{candidate} for the volume operator. At this point one can forget the construction, study the properties of this operator both in the deep quantum regime and in the semiclassical regime and understand if it actually has the meaning of ``volume of a region'' at both levels.

Given the number of choices to be made, it is not surprising that two distinct mathematically well-defined volume operators exist in the literature, one due to Rovelli and Smolin \cite{Rovelli:1994ge} the other to Ashtekar and Lewandowski \cite{Ashtekar:1997fb}. Both of them act non-trivially only at the nodes of a spin~network state. In this sense, both of them fit into the picture discussed in section \ref{sec:picture}. For a discussion of the relation between the two operators see \cite{Ashtekar:1997fb,Fairbairn:2004qe,Giesel:2005bk,Giesel:2005bm}. Here we describe in detail some aspects of the Rovelli-Smolin construction \cite{Rovelli:1994ge} of the volume operator as it will play a role in the following.

\subsubsection{External regularization of the volume}\label{sec:external V}
The construction of the regularized expression for the volume to be used as starting point for quantization goes through the following steps\footnote{The construction we discuss here is based on \cite{Rovelli:1994ge}, \cite{DePietri:1996pj}, \cite{Rovelli:2004tv} but does not completely coincide with it. See also \cite{Lewandowski:1996gk} and   \cite{Ashtekar:1997fb} for a comprehensive discussion of the many subtleties involved and a comparison with the Ashtekar-Lewandowski construction.}:
\begin{itemize}
	\item[i)] The integral over $R$ is replaced by the limit of a Riemann sum. More specifically, we choose coordinates $x^a$ in a neighbourhood in $\Sigma$ containing $R$ and consider a partition of the neighbourhood in cubic cells $R_I$ of coordinate side $\Delta x$. Therefore the region $R$ is contained in the union of a number of cells, $R\subseteq \cup_I R_I$, and the integral $\int_R d^3x$ can be approximated from above by the sum $\sum_I (\Delta x)^3$ with $I$ running on the cells containing points of $R$.
	\item[ii)] The argument of the square root in (\ref{eq:vol}) in a point contained in the cell $R_I$ is written in terms of the limit of a quantity $W_{\Delta x}(x_I)$  given by a triple surface integral over the boundary of the cell:
\begin{align}
W_{\Delta x}(x_I)=&\frac{1}{8\times 3!} \frac{1}{(\Delta x)^6} \int_{\p R_I}d^2\sigma \int_{\p R_I}d^2\sigma' \int_{\p R_I}d^2\sigma''\;\times\\ 
&\times|T_{x_I}^{ijk}(\sigma,\sigma',\sigma'')\,E^a_i(\sigma)\,n_a(\sigma)\;E^b_j(\sigma')\,n_b(\sigma')\,E^c_k(\sigma'')\,n_c(\sigma'')\,|
	\label{eq:W delta x}
\end{align}
In (\ref{eq:W delta x}) the following notation has been used. Let's consider a surface $S$, a choice of local coordinates $\sigma^\alpha$ 
and an embedding of $S$ in $\Sigma$ given by $x^a=X^a(\sigma)$. The quantity $n_a(\sigma)$ is defined as
\begin{equation}
n_a(\sigma)=\eps_{abc}\frac{\p X^b}{\p \sigma^1}\frac{\p X^c}{\p \sigma^2}\;.
	\label{eq:n_a}
\end{equation}
Notice that in (\ref{eq:W delta x}) we are considering a surface given by the boundary of a cubic cell, therefore the function $n_a(\sigma)$ is not continuous in $\sigma$. By $E^a_i(\sigma)$ we simply mean $E^a_i(X(\sigma))$. The function $T_{x_I}^{ijk}(\sigma,\sigma',\sigma'')$ has been inserted in order to guarantee the $SU(2)$-gauge invariance of the non-local expression (\ref{eq:W delta x}). It is given by 
\begin{equation}
T_{x_I}^{ijk}(\sigma,\sigma',\sigma'')=\eps^{i'j'k'}\; \mc{D}^{(1)}(h_{\gamma^1_{x_I\sigma}}[A])_{i'}^{\phantom{i'}i}\; \mc{D}^{(1)}(h_{\gamma^2_{x_I\sigma'}}[A])_{j'}^{\phantom{j'}j}\; \mc{D}^{(1)}(h_{\gamma^3_{x_I\sigma''}}[A])_{k'}^{\phantom{k'}k}	
	\label{eq:Tijk}
\end{equation}
where $\gamma^1_{x_I\sigma}$, $\gamma^2_{x_I\sigma'}$ and $\gamma^3_{x_I\sigma''}$ are three curves embedded in $R_I$ having starting point $x_I$ in $R_I$ and ending at a point on the boundary of $R_I$ given by $X(\sigma)$, $X(\sigma')$ and $X(\sigma'')$ respectively. As already explained, by $\mc{D}^{(1)}(h_\gamma[A])_i^{\phantom{i}j}$ we mean the holonomy of the real $SU(2)$ connection along the curve $\gamma$, taken in the adjoint representation. 

In the limit $\Delta x \to 0$, under the assumption of smooth $E^a_i(x)$ and $A_a^i(x)$, we have that\footnote{In order to show it, it is useful to recall the formula 
\begin{equation*}
\eps^{i'j'k'}	E^{a'}_{i'} E^{b'}_{j'} E^{c'}_{k'}= \frac{1}{3!}\big(\eps_{abc} \eps^{ijk} E^a_i E^b_j E^c_k\big)\; \eps^{a'b'c'}\;.
\end{equation*}} $W_{\Delta x}(x_I)$ goes to $\frac{1}{3!}|\eps^{ijk}\eps_{abc}E^a_i(x_I) E^b_j(x_I) E^c_k(x_I)|$. Therefore we have
\begin{equation}
	V(R)=\lim_{\Delta x \to 0} \sum_I (\Delta x)^3 \sqrt{W_{\Delta x}(x_I)}\;.
	\label{eq:V=lim Dx3}
\end{equation}
Notice that the factor $(\Delta x)^{-6}$ present in $W_{\Delta x}(x_I)$ cancels with the $(\Delta x)^3$ appearing in (\ref{eq:V=lim Dx3}). This corresponds to the fact that $\sqrt{\frac{1}{3!}|\eps^{ijk}\eps_{abc}E^a_i E^b_j E^c_k|}$ is a density of weight one and can be integrated with the measure $\int d^3 x$. As a result, in (\ref{eq:V=lim Dx3}) $\Delta x$ appears only implicitly in the definition of the surface $\p R_I$.
   \item[iii)] The surface $\p R_I$ can be partitioned in square cells $S_I^\alpha$ so that $\p R_I=\cup_\alpha S_I^\alpha$. As a result the triple integral over $\p R_I$ can be replaced by a triple Riemann sum. In this way we end up with an expression depending only on fluxes and holonomies. Defining the quantity $Q_{I\alpha\beta\gamma}$ for a cell $R_I$ and three surfaces $S_I^\alpha$, $S_I^\beta$ and $S_I^\gamma$ as 
\begin{equation}
Q_{I\alpha\beta\gamma}=T_{x_I}^{ijk} \, F_i(S_I^\alpha)\,F_j(S_I^\beta)\,F_k(S_I^\gamma)\;,	
	\label{eq:Q}
\end{equation}
we have\footnote{The prime in the sum in (\ref{eq:regularized V_I}) stands for sum restricted to distinct $\alpha,\beta,\gamma$. This corresponds to a point-splitting of the integral over $(\p R_I)^3$.}
\begin{equation}
	V_I= \sqrt{\frac{1}{8\times 3!} {\sum}'_{\alpha\,\beta\,\gamma}|Q_{I\alpha\beta\gamma}|}
	\label{eq:regularized V_I}  
\end{equation}
and
\begin{equation}
	  V(R)=\lim_{\Delta x \to 0} \sum_I V_I\;.
	\label{eq:regularized V}
\end{equation}
Notice that while the regularized expression depends both on the $E^a_i$ and on $A_a^i$, the limit depends only on the electric field. 
\end{itemize}
Step (ii) and (iii) can be called a \emph{fluxization} of the Riemann sum of step (1).

\subsubsection{Quantization of the volume}\label{sec:quantization V}
Having constructed a sequence of regularized expressions having the appropriate classical limit, we can now attempt to promote (\ref{eq:regularized V}) to a quantum operator by invoking the known action of the holonomy and of the flux on cylindrical functions, namely
\begin{equation}
	\widehat{V(R)}\, \Psi_{\Gamma,f}[A]=\lim_{\Delta x \to 0}\,\Big(\sum_I \widehat{V}_I \,\Psi_{\Gamma,f}[A]\Big)\;.
	\label{eq:V on cyl}
\end{equation}
To be more specific, we need to define a consistent family of operators for finite $\Delta x$ and given cylindrical function. This step requires a number of choices which we state below. Then we can analyze the existence of the limit in the operator topology.

Let $\Gamma$ be a closed graph embedded in $\Sigma$ and made of $N$ nodes connected by $M$ links $\{e_1,..,e_M\}$. A $SU(2)$-gauge invariant state which is cylindrical with respect to the graph $\Gamma$ is defined as 
\begin{equation}
	\Psi_{\Gamma,f}[A]=f(h_{e_1}[A],..h_{e_M}[A])
	\label{eq:cyl}
\end{equation}
with $f$ a class function on $SU(2)^M$. In order to define the regularized operator $\hat{V}_I$ for finite $\Delta x$, an adaptation of the partition of $R$ to the graph $\Gamma$ is needed. The partition of the region $R$ in cells $R_I$ is refined so that 
\begin{itemize}
	\item[-] nodes of $\Gamma$ can fall only in the interior of cells;
	\item[-] a cell $R_I$ contains at most one node. In case it contains no node, then it can contain at most one link;
	\item[-] the boundary $\p R_I$ of a cell intersects a link exactly once if the link ends up at a node contained in the cell and exactly twice if it does not.
\end{itemize}
Moreover we assume that the partition of the surfaces $\p R_I$ in cells $S_I^\alpha$ is refined so that links of $\Gamma$ can intersect a cell $S_I^\alpha$ only in its interior and each cell $S_I^\alpha$ is punctured at most by one link.

Next we focus on the action of the operator $\hat{Q}_{I\alpha\beta\gamma}$ obtained quantizing canonically expression (\ref{eq:Q}):
\begin{equation}
\hat{Q}_{I\alpha\beta\gamma}=\hat{T}_{x_I}^{ijk} \, \widehat{F}_i(S_I^\alpha)\,\widehat{F}_j(S_I^\beta)\,\widehat{F}_k(S_I^\gamma)\;.	
	\label{eq:Q quantum}
\end{equation}
Let's call it the three-hand operator. Notice that we don't need a specific ordering of the fluxes and the holonomies thanks to the fact that the self-grasping vanishes\footnote{This is a straightforward consequence of the fact that $\delta^{il} (T^{(1)}_i)^k_{\phantom{k}l}=0$.}. From properties of the action of the flux operator on a holonomy, we know that when the operator $\hat{Q}_{I\alpha\beta\gamma}$ acts on a state $\Psi_{\Gamma,f}[A]$ the result is zero unless each of the surfaces $S_I^\alpha$, $S_I^\beta$ and $S_I^\gamma$ is punctured by a link of $\Gamma$. As a result if the cell $R_I$ does not contain nodes of $\Gamma$, then $\hat{Q}_{I\alpha\beta\gamma}$ annihilates the state.



\begin{figure}[t]
\centering
\begin{minipage}[b]{0.32\textwidth}
\begin{tabular}{c}
\includegraphics[height=.7\textwidth]{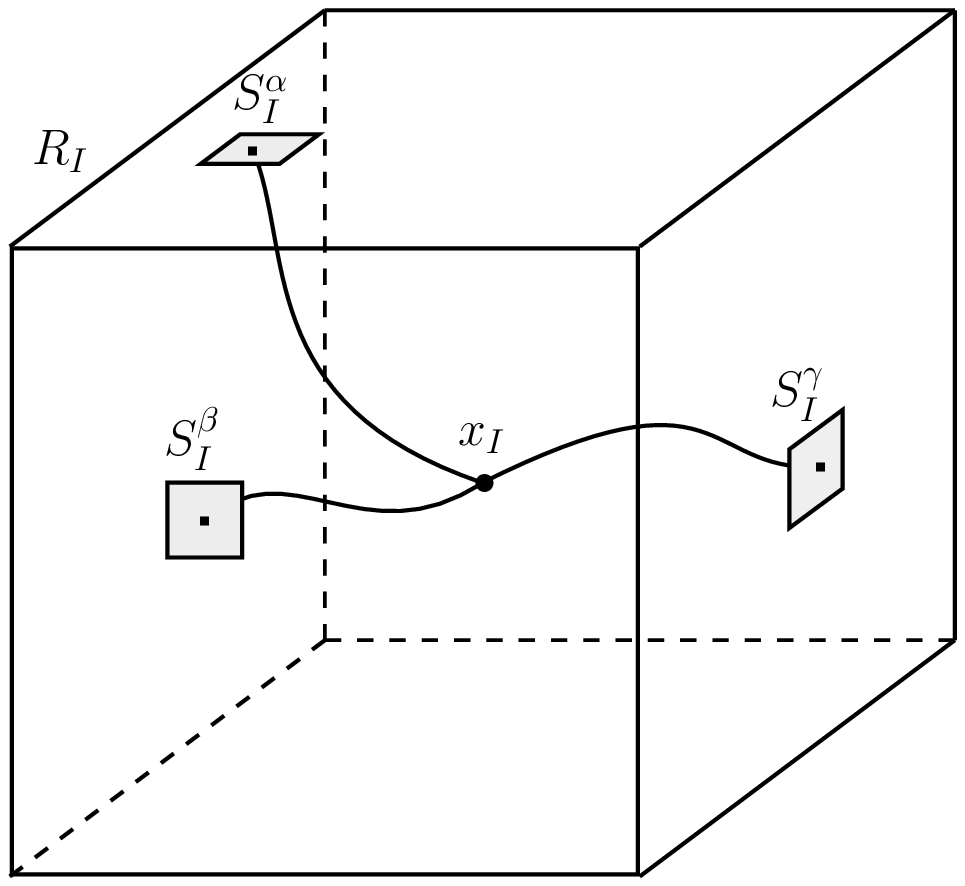}\\
(a)
\end{tabular}
\end{minipage}
\begin{minipage}[b]{0.32\textwidth}
\begin{tabular}{c}
\includegraphics[height=.7\textwidth]{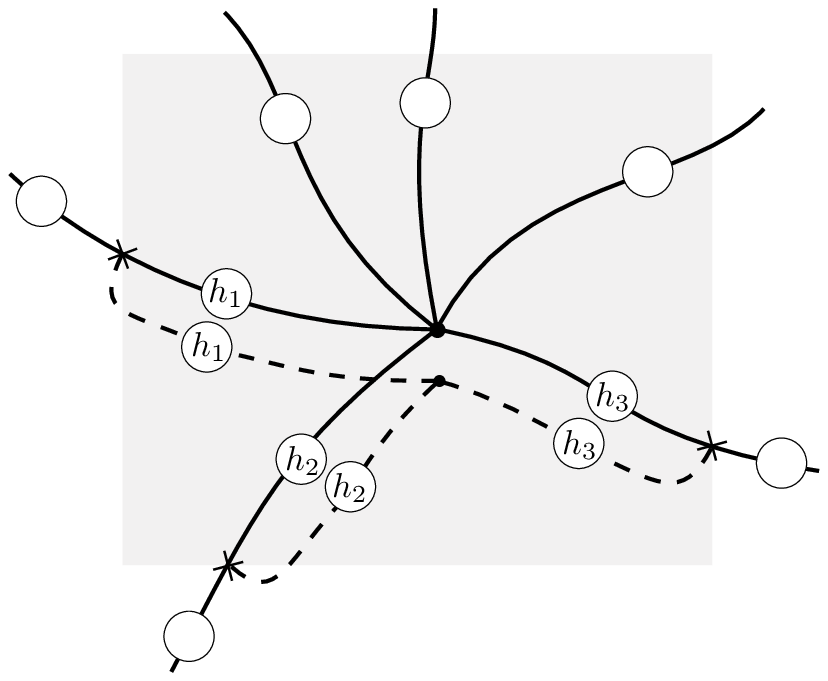}\\
(b)
\end{tabular}
\end{minipage}
\begin{minipage}[b]{0.32\textwidth}
\begin{tabular}{c}
\includegraphics[height=.7\textwidth]{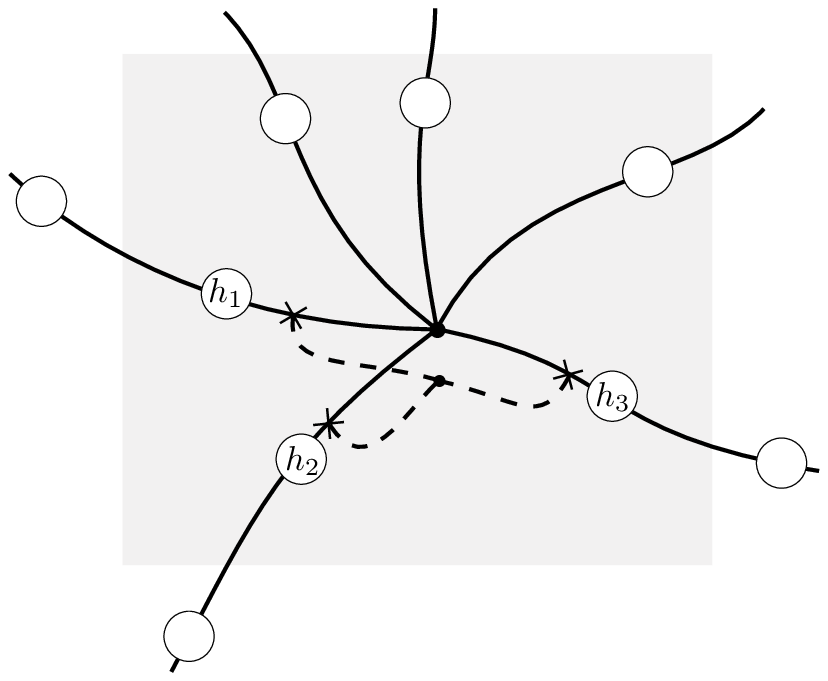}\\
(c)
\end{tabular}
\end{minipage}
\caption{(a) A cubic cell with the regularized quantity (\ref{eq:Q}) represented. (b) Action of the three-hand operator. The cubic cell is shown in gray. (c) Shrinking property of the three-hand operator.}
\label{fig:volume}
\end{figure}

Now let's focus on a cell $R_I$ which contains a node of $\Gamma$. In this case, some further adaptation of the regularized expression (\ref{eq:Q}) to the graph $\Gamma$ is required. The point $x_I$ and the three curves introduced by $T_{x_I}^{ijk}$ in the definition of the regularized volume are adapted to the graph $\Gamma$ in the following way:
\begin{itemize}
	\item[-] the point $x_I$ in (\ref{eq:Tijk}) is chosen to coincide with the position of the node,
	\item[-] the three curves $\gamma^1_{x_I\sigma}$, $\gamma^2_{x_I\sigma}$ and $\gamma^3_{x_I\sigma}$ are adapted to three of the links of $\Gamma$ originating at the node contained in the cell $R_I$.
\end{itemize}
As a result the appropriate labels for the operator (\ref{eq:Q quantum}) are a node $n$ and a triple of links $e_1,e_2,e_3$. We have that, when the operator acts on a state of the spin~network basis of $\mc{K}_0(\Gamma)$, its action is the following
\begin{align}
	&\hat{Q}_{n\, e_1 e_2 e_3}\;\Psi_{\Gamma,j,i_k}[A]\,=\,
		\hat{Q}_{n\, e_1 e_2 e_3}\, \Big(\mc{D}^{(j_1)}(h_{e_1}[A])_{m_1'}^{\ph{m_1}m_1}\cdot\cdot\, \mc{D}^{(j_L)}(h_{e_L}[A])_{m_L'}^{\ph{m_L}m_L}\;{v_k}^{(j_1\cdot\cdot j_L)}_{m_1\cdot\cdot m_L}\Big)\,\times \textrm{rest}^{m'_1\cdot\cdot m'_L}= \nonumber\\
		&\hspace{.5 em}=\,(8\pi \gamma L_P^2)^3 \;\eps^{i'j'k'} \mc{D}^{(1)}(h_{e_1}[A])_{i'}^{\ph{i}i} \mc{D}^{(1)}(h_{e_2}[A])_{j'}^{\ph{j}j} \mc{D}^{(1)}(h_{e_3}[A])_{k'}^{\ph{k}k}\;		
		(T^{(j_1)}_i)_{m_1'}^{\ph{m}m_1''} (T^{(j_2)}_j)_{m_2'}^{\ph{m}m_2''} (T^{(j_3)}_k)_{m_3'}^{\ph{m}m_3''} \times\nonumber\\		
&\hspace{2em}\times	\Big(\mc{D}^{(j_1)}(h_{e_1}[A])_{m_1''}^{\ph{m_1}m_1}\cdot\cdot\, \mc{D}^{(j_L)}(h_{e_L}[A])_{m_L'}^{\ph{m_L}m_L}\;{v_k}^{(j_1\cdot\cdot j_L)}_{m_1\cdot\cdot m_L}\Big)\,\times \textrm{rest}^{m'_1\cdot\cdot m'_L}
\end{align}
This expression has the diagrammatic representation \ref{fig:volume}-(b).


%
 
The adaptation of $T_{x_I}^{ijk}$ to the graph $\Gamma$  as described above has the following remarkable property: shrinking the region $R_I$ corresponds to moving the graspings in figure \ref{fig:volume}-(b) towards the node; however, thanks to the invariance properties of the intertwiner inserted by the grasping, the result of the triple-grasping is independent of the position of the grasping and it can be moved to the node as shown in figure \ref{fig:volume}-(c). In formulae we have the identity
\begin{align*}
&\eps^{i'j'k'} \mc{D}^{(1)}(h_{e_1}[A])_{i'}^{\ph{i}i} \mc{D}^{(1)}(h_{e_2}[A])_{j'}^{\ph{j}j} \mc{D}^{(1)}(h_{e_3}[A])_{k'}^{\ph{k}k}\;		
		(T^{(j_1)}_i)_{m_1'}^{\ph{m}m_1''} (T^{(j_2)}_j)_{m_2'}^{\ph{m}m_2''} (T^{(j_3)}_k)_{m_3'}^{\ph{m}m_3''} \times\\		
&\hspace{15em}\times	\Big(\mc{D}^{(j_1)}(h_{e_1}[A])_{m_1''}^{\ph{m_1}m_1}\cdot\cdot\, \mc{D}^{(j_L)}(h_{e_L}[A])_{m_L'}^{\ph{m_L}m_L}\;{v_k}^{(j_1\cdot\cdot j_L)}_{m_1\cdot\cdot m_L}\Big)=\\
&	= \mc{D}^{(j_1)}(h_{e_1}[A])_{m_1'}^{\ph{m_1}m_1}\cdot\cdot\, \mc{D}^{(j_L)}(h_{e_L}[A])_{m_L'}^{\ph{m_L}m_L}\;\Big(\eps^{ijk} 	
		(T^{(j_1)}_i)_{m_1}^{\ph{m}m_1''} (T^{(j_2)}_j)_{m_2}^{\ph{m}m_2''} (T^{(j_3)}_k)_{m_3}^{\ph{m}m_3''} {v_k}^{(j_1\cdot\cdot j_L)}_{m_1''\cdot\cdot m_L}\Big)
\end{align*}
where the left hand side corresponds to the evaluation of the diagram \ref{fig:volume}-(b) while the right hand side to the evaluation of \ref{fig:volume}-(c).

As a result we have that, for finite $\Delta x$ and with the refinement of the partition and adaptations to the graph described above, the action of the operator $\hat{Q}_{I\alpha\beta\gamma}$ on a spin~network node is
\begin{itemize}
	\item independent of $\Delta x$,
	\item does not change the graph of the state,
	\item does not change the spin labelling of the state.
\end{itemize}
Therefore its matrix elements are non-trivial only in the intertwiner sector and can be computed using standard recoupling techniques \cite{DePietri:1996pj,Thiemann:1996au,Brunnemann:2004xi}. In formulae we have that
\begin{equation}
	\hat{Q}_{n\, e_1 e_2 e_3}\;\Psi_{\Gamma,j,i_k}[A]\,=\,\sum_h (Q_{n\, e_1 e_2 e_3})_k^{\ph{k}h}\;\;\Psi_{\Gamma,j,i_h}[A]
	\label{eq:Q on intertwiner}
\end{equation}
Given the valence of the node and the spins of the incoming links, we have a finite dimensional hermitian matrix\footnote{In the case of a trivalent node, the intertwiner space is one-dimensional. Therefore we have that trivalent nodes are always eigenstates of the operator $Q_I$ (and of the operator $V_I$). Explicit computation \cite{Loll:1995tp} shows that the eigenvalue is zero. As a result the simplest non-trivial case is for $4$-valent nodes.} $(Q_{n\, e_1 e_2 e_3})_k^{\ph{k}h}$. The operator $\hat{V}_I$ involves taking a modulus of such matrix and a square root of a sum of matrices and this can be done through spectral decomposition. This defines the operator $\hat{V}_I$ for finite $\Delta x$. Moreover this is enough to define the action of the operator $\widehat{V}(R)$ on a given spin~network state too as, once an appropriate refinement of the partition is reached, the action of the regularized operator is independent of $\Delta x$ and the limit in equation (\ref{eq:regularized V}) is guaranteed to exist as it is simply the limit of a constant. Having defined the matrix elements of the operator $\widehat{V}(R)$ on a orthonormal basis, the spin~network basis, then one can attempt to promote it to a well-defined operator on the whole Hilbert space $\mc{K}_0$ through self-adjoint extension.

The volume operator for a region has the remarkable feature that it can be expressed in terms of `elementary' volume operators. Let's consider a graph $\Gamma$ embedded in $\Sigma$, focus on a node $n$ of $\Gamma$ and choose a region $R_n$ such that it contains the node $n$, but does not contain any other node of $\Gamma$. We call the region $R_n$ \emph{dual} to the node $n$. Then we consider the Hilbert space $\mc{K}_0(\Gamma)$ spanned by spin~network states having exactly $\Gamma$ as graph. This is a subspace of the Loop Quantum Gravity state space $\mc{K}_0$. On this Hilbert space the operator $\widehat{V}(R_n)$ is well defined, acts only on the intertwiner space at the node $n$ and the matrix elements do not depend on the specific choice\footnote{That is, if two regions  $R_n$ and $R'_n$ are both dual to the node $n$, then the operators $\widehat{V}(R_n)$ and $\widehat{V}(R'_n)$ coincide on $\mc{K}_0(\Gamma)$, i.e. they have the same matrix elements on the spin~network basis of $\mc{K}_0(\Gamma)$.} of the surface $R_n$. As a result it can be said that it measures the volume of a region dual to the node $n$. We call this operator the `elementary' volume operator for the node $n$ and indicate it as $\hat{V}_n$. For a generic region $R$ the volume operator on $\mc{K}_0(\Gamma)$ is given by a sum of `elementary' volume operators
\begin{equation}
	\widehat{V(R)}\,\Psi_{\Gamma, f}[A]=\sum_{n\subset R} \hat{V}_n \Psi_{\Gamma, f}[A]\;.
	\label{eq:elementary volume}
\end{equation}
This property enlightens the quantum geometrical meaning of states belonging to $\mc{K}_0(\Gamma)$, and in particular of spin~network states. Moreover it offers the possibility of identifying a region $R$ in a relational way, i.e. with respect to the state of the gravitational field \cite{Rovelli:2001bz}.

This concludes our description of the construction of the Rovelli-Smolin volume operator. It is in many ways over-detailed and nevertheless certainly not complete. However we want to stress again that the aim of the construction described above is to provide motivation and a guide in the choice of an operator, not a ``proof'' that the constructed quantum operator corresponds to the classical volume. The other way around. It is analyzing its spectrum and its eigenstates that the geometrical meaning of ``volume of a region of space'' can be established in the deep quantum regime. See \cite{DePietri:1996pj,Major:2001zg} and \cite{Thiemann:1996au,Brunnemann:2004xi,Brunnemann:2007ca,Brunneman:2007as}. As far as the semiclassical behaviour is involved, things are less clear due to lack of control on the semiclassical states of the theory both at the kinematical and at the dynamical level. Having quantized a classical expression a priori guaranties only that the naive semiclassical limit $\hbar \to 0$ is the appropriate one. In section \ref{sec:semiclassical} of this paper we discuss a good semiclassical property of the Rovelli-Smolin volume operator on certain specific superpositions of spin~network states.

\vspace{2em}

In next section we focus on the construction of the length operator, identify the difficulties one encounters trying to apply the quantization procedure described above for the volume, discuss our strategy and implement it in some detail. 

\section{Construction of the length operator}\label{sec:construction}


Again the starting point is a classical expression, the expression for the length of a curve. Given a curve $\gamma$ embedded in the $3$-manifold $\Sigma$,
\begin{equation}
\begin{array}{rccc}
\gamma:&[0,1] &  \to    & \Sigma   \\[1ex]
       &	s   & \mapsto & \gamma^a(s)
\end{array}	\label{eq:gamma}
\end{equation}
the length $L(\gamma)$ of the curve is a functional of the electric field $E^a_i$ given by the following one-dimensional integral
\begin{equation}
L(\gamma)=\int_0^1\!\!ds\;\sqrt{\de_{ij}\, G^i(s)\,G^j(s)}\;,
	\label{eq:classical L}
\end{equation}
where
\begin{equation}	G^i(s)=\frac{\frac{1}{2}\eps^{ijk}\eps_{abc}E^b_j\, E^c_k \;\dot{\gamma}^a(s)}{\sqrt{\frac{1}{3!}|\eps^{ijk}\eps_{abc}E^a_i E^b_j E^c_k|}}\;.
	\label{eq:classical G}
\end{equation}
In (\ref{eq:classical G}) all the $E^a_i(x)$ are evaluated at $x^a=\gamma^a(s)$ and $\dot{\gamma}^a(s)=\frac{d}{ds}\gamma^a(s)$. The quantity (\ref{eq:classical G}) corresponds to the pullback of the triad on the curve, $G^i(s)=e_a^i(\gamma(s))\,\dot{\gamma}^a(s)\,$.

Given the classical expression (\ref{eq:classical L})--(\ref{eq:classical G}) for the length, then one can write the one-dimensional integral as a Riemann sum over cells of coordinate size $\Delta s$ as done for the volume in step (i) of section \ref{sec:external V}, find a fluxization of this expression as in steps (ii) and (iii) of section \ref{sec:external V} and attempt to promote to an operator the regularized expression.


\subsection{Difficulties in quantizing the length}\label{sec:difficulties}
The classical expression of the length is a one-dimensional integral of a rather complicated, non-polynomial function of the electric field. As a result one has to expect to encounter the following two kinds of difficulties in trying to quantize it using the procedure described above for the volume operator: 
\begin{itemize}
	\item[a)] the fluxization procedure described by steps (ii) and (iii) in section \ref{sec:external V} would correspond to finding a way to write a one-dimensional object (the summand in a Riemann sum) in terms of a function of two-dimensional objects, namely the fluxes. It is far from clear that this can be done at all. Assuming to find one such fluxization, then one has to promote it to an operator acting on cylindrical functions with given graph. This would require an adaptation to the graph both of the partition and of the curves corresponding to the holonomies appearing in a regularized expression for the length. As always a number of choices has to be made. For instance one could quantize the denominator in (\ref{eq:classical G}) has done for the volume. However, as far as the numerator (\ref{eq:classical G}) is concerned a specific choice is needed, namely one which knows the direction $\dot{\gamma}^a$ of the curve. Without a strategy which allows to take this into account, one will almost certainly end up with an operator which has nothing to do with the length of the curve $\gamma$; 
  \item[b)] the denominator in (\ref{eq:classical G}) is the local volume density. Therefore one should expect that the definition of the length operator involves the inverse of the local volume operator $\hat{V}_I$ defined previously. However the operator $\hat{V}_I$ is known to be non-invertible as it has a huge kernel.
\end{itemize}

\subsection{The strategy}\label{sec:strategy}
The strategy we follow in this paper is to focus on difficulty (a): we take seriously the dual picture of quantum geometry described in section \ref{sec:picture} and use it as a guide for constructing the length operator. The external regularization described for the volume in section \ref{sec:external V} turns out to be an extremely well-suited tool. In the next section we will show how to fluxize the classical expression (\ref{eq:classical L}) and how to adapt it to the graph of a cylindrical function. Our fluxization procedure rests on the fact that a curve can be identified as the intersection of two surfaces. As a result, the tangent to the curve can be written in terms of the normals to the two surfaces. 

As far as difficulty (b) is concerned, we consider it as technical and we deal with it as such. At the end of our analysis, studying the semiclassical behaviour of the length operator, we will find evidence that (b) is not a separate issue and is already solved when (a) is appropriately taken into account.

\subsection{Construction of the length operator}\label{sec:construction subsec}

\subsubsection{External regularization of the length of a curve}
\begin{figure}[t]
\centering
\begin{minipage}[b]{0.32\textwidth}
\begin{tabular}{c}
\includegraphics[height=.9\textwidth]{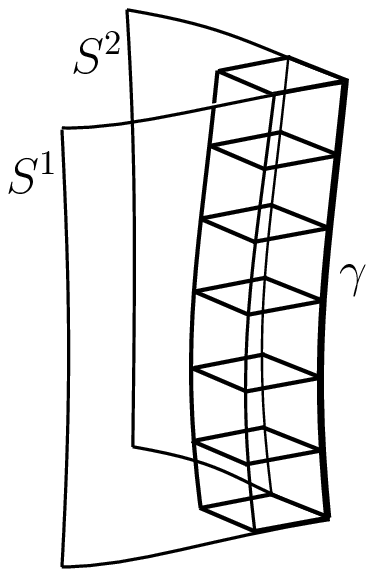}\\
(a)
\end{tabular}
\end{minipage}
\begin{minipage}[b]{0.32\textwidth}
\begin{tabular}{c}
\includegraphics[height=.7\textwidth]{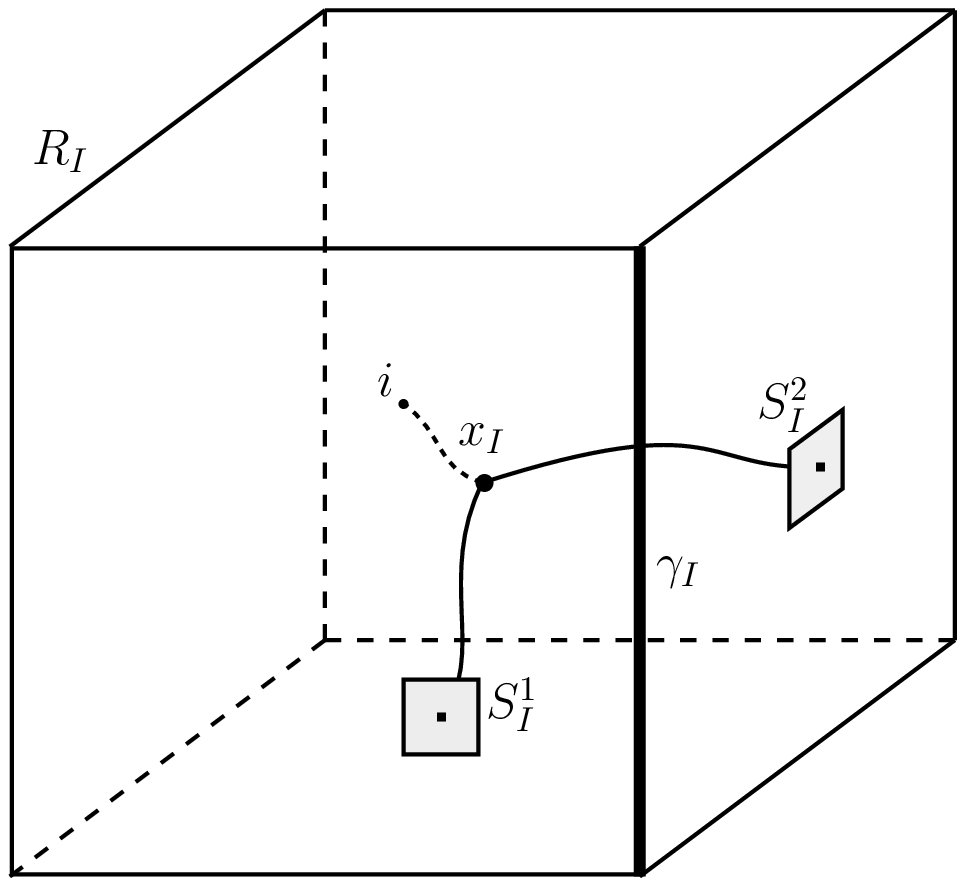}\\
(b)
\end{tabular}
\end{minipage}
\begin{minipage}[b]{0.32\textwidth}
\begin{tabular}{c}
\includegraphics[height=.7\textwidth]{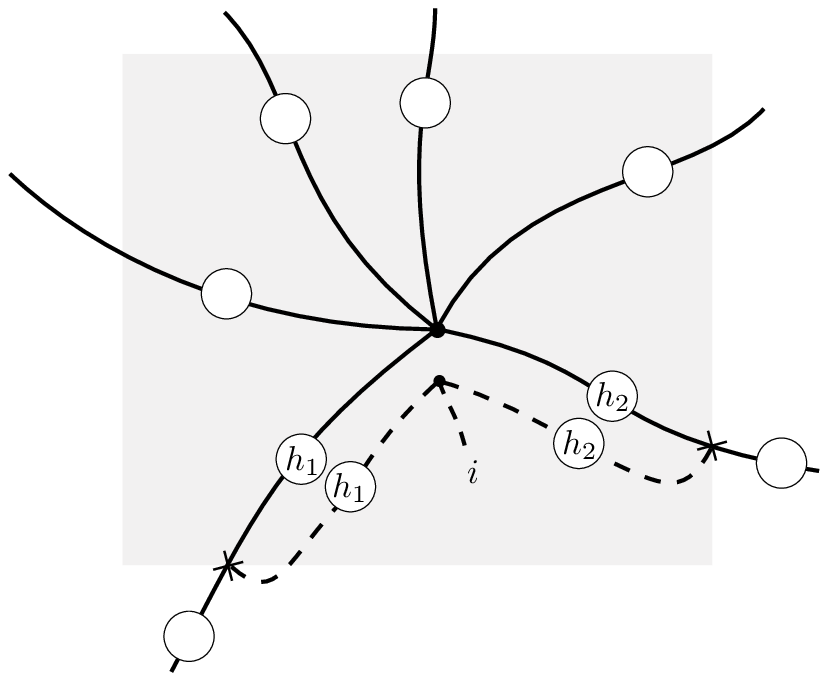}\\
(c)
\end{tabular}
\end{minipage}
\caption{(a) A curve as the intersection of two surfaces. The decomposition in cubic cells is shown. (b) Cubic cell with the regularized quantity (\ref{eq:regularized Y}) shown. (c) Action of the two-hand operator on a spin~network state. The cubic cell is shown in gray.}
\label{fig: regularization L}
\end{figure}

In this section we present an \emph{external} regularization of the length of a curve and use it as starting point for quantization. The regularization we consider here goes through the following steps:
\begin{itemize}
	\item[i)] The one-dimensional integral in (\ref{eq:classical L}) is replaced by the limit of a Riemann sum. For instance, we consider a partition of the curve $\gamma$ in segments $\gamma_I$ corresponding to embeddings $x^a=\gamma^a(s)$ with $s$ in the interval $[I\, \Delta s, (I+1)\Delta s]$ 
	so that $\gamma=\cup_I \gamma_I$. Then we can write (\ref{eq:classical L}) as
\begin{equation}
	L(\gamma)=\lim_{\Delta s \to 0} \sum_I \Delta s\;\sqrt{\de_{ij} G_{\Delta s}^i(s_I) G_{\Delta s}^j(s_I)}
	\label{eq:length Riemann sum}
\end{equation}
with $s_I$ a point belonging to $[I\, \Delta s, (I+1)\Delta s]$. The subscript $\Delta s$ in $G_{\Delta s}^i(s_I)$ is there to recall that the regularized expression for $G^i(s)$ that we are going to introduce below depends on the step $\Delta s$.
  
\vspace{1ex}
  
  Given the curve $\gamma$ embedded in $\Sigma$, let's consider two surfaces $S_1$ and $S_2$ which intersect at $\gamma$. A way to visualize it is to choose coordinates $x^a=(\sigma_1,\sigma_2,s)$ in $\Sigma$ such that the curve has embedding $x^a=\gamma^a(s)=(0,0,s)$, the surface $S_1$ is the $\sigma_2=0$ surface and the surface $S_2$ is the $\sigma_1=0$ surface. 
  Then we can consider a partition of $\Sigma$ in cells which are cubic with respect to the coordinates $x^a=(\sigma_1,\sigma_2,s)$ and have coordinate size $\Delta s$. See figure \ref{fig: regularization L}-(a) for reference. As a result, we have that the cell $R_I=\{x^a\in \Sigma \,|\; \sigma_1\in[0,\Delta s],\sigma_2\in[0,\Delta s], s \in [I\, \Delta s, (I+1)\Delta s] \}$ has the segment $\gamma_I$ as side. Moreover the segment $\gamma_I$ corresponds to the intersection of the two surfaces $S^1_I$ and $S^2_I$ belonging to the boundary of $R_I$.  
  \item[ii)] Thanks to the partition introduced above, now we can make the first step of the fluxization procedure, namely we can write $G_i(s)$ in terms of surface integrals:
\begin{align}
	&\hspace{-1em}G_{\Delta s}^i(s_I)=\label{eq:surface int G}\\
	&=\frac{\frac{1}{2}\frac{1}{(\Delta s)^4}\int_{S_I^1}d^2\sigma\int_{S_I^2}d^2\sigma'\;V_{x_I}^{ijk}(\sigma,\sigma') E^a_j(\sigma) n_a(\sigma) E^b_k(\sigma') n_b(\sigma')}{\sqrt{\frac{1}{8\times 3!} \frac{1}{(\Delta s)^6} \int\limits_{\p R_I}\hspace{-.6em}d^2\sigma\hspace{-.6em} \int\limits_{\p R_I}\hspace{-.6em}d^2\sigma'\hspace{-.6em} \int\limits_{\p R_I}\hspace{-.6em}d^2\sigma''|T_{x_I}^{ijk}(\sigma,\sigma',\sigma'')\,E^a_i(\sigma)\,n_a(\sigma)\;E^b_j(\sigma')\,n_b(\sigma')\,E^c_k(\sigma'')\,n_c(\sigma'')\,|}}
	\nonumber
\end{align}
For the notation used we refer to step (ii) of section \ref{sec:external V}. The double surface integral at the numerator corresponds to the numerator in the right hand side of (\ref{eq:classical G}). The quantity $V_{x_I}^{ijk}(\sigma,\sigma')$ has been introduced in order to make $SU(2)$-gauge invariant such non-local expression in the field $E^a_i(x)$. It is defined as
\begin{equation}
V_{x_I}^{ijk}(\sigma,\sigma')=\eps^{i\,j'k'}\;  \mc{D}^{(1)}(h_{\gamma^1_{x_I\sigma'}}[A])_{j'}^{\phantom{j'}j}\; \mc{D}^{(1)}(h_{\gamma^2_{x_I\sigma''}}[A])_{k'}^{\phantom{k'}k}	\;.
	\label{eq:Vijk}
\end{equation}
In the limit $\Delta s \to 0$ the numerator goes to $\frac{1}{2}\eps^{ijk}\eps_{abc}E^b_j\, E^c_k \;\dot{\gamma}^a(s)$ thanks to the fact that
\begin{equation}
 \lim_{\sigma_1, \sigma_2\to 0} \eps^{abc} n_b(\sigma_1,s) n_c(\sigma_2,s)= \dot{\gamma}^a(s)
	\label{eq:wedge normals}
\end{equation}

As far as the denominator is concerned, it corresponds to the external regularization of the volume density discussed in step (ii) of section \ref{sec:external V}.
  \item[iii)] now we make the second step of the fluxization procedure, namely we write the surface integrals in (\ref{eq:surface int G}) as Riemann sums of fluxes. The surface $\p R_I$ is partitioned into square cells $S_{I\alpha}$ so that $\p R_I=\cup_\alpha S_{I\alpha}$. In particular we have $S_I^1=\cup_\alpha S_{I\alpha}^1$ and $S_I^2=\cup_\alpha S_{I\alpha}^2$ for the two surfaces $S_I^1$ and $S_I^2$. As a result we have that $G_{\Delta s}^i(s_I)$ of (\ref{eq:surface int G}) can be written as the limit of $(\Delta s)^{-1}$ times a quantity $G^i_I$ defined as follows 
\begin{equation}
	G^i_I=\frac{\frac{1}{2}\sum_{\alpha\,\beta} Y_{I\alpha\beta}^i}{\sqrt{\frac{1}{8\times 3!} \sum'_{\alpha\,\beta\,\gamma}|Q_{I\alpha\beta\gamma}|}}
	\label{eq:regularized G}
\end{equation}
where $Y_{I\alpha\beta}^i$ is given by
\begin{equation}
	Y_{I\alpha\beta}^i=V_{x_I}^{ijk} F_j(S_{I\alpha}^1) F_k(S_{I\beta}^2)
	\label{eq:regularized Y}
\end{equation}
and $Q_{I\alpha\beta\gamma}$ has been introduced in (\ref{eq:Q}) (notice that the denominator in equation (\ref{eq:regularized G}) is the regularized local volume $V_I$ introduced in (\ref{eq:regularized V_I})). Then the length of the segment $\gamma_I$ can be defined in terms of $G^i_I$ introduced above and is given by
\begin{equation}
	L_I=\sqrt{\de_{ij}\; G^i_I \, G^j_I}\,.
	\label{eq:regularized L_I}
\end{equation}
As a result we have that the length of the curve $\gamma$ can be written as the limit $\Delta s \to 0$ of a sum of terms depending on $\Delta s$ only implicitly
\begin{equation}
	L(\gamma)=\lim_{\Delta s \to 0} \sum_I L_I\,.
	\label{eq:regularized L}
\end{equation}
\end{itemize}

\subsubsection{Quantization of the regularized expression}
Having constructed a sequence of regularized expressions having the appropriate classical limit, we can now attempt to promote (\ref{eq:regularized L}) to a quantum operator by invoking the known action of the holonomy and of the flux on cylindrical functions, namely
\begin{equation}
	\widehat{L}(\gamma)\, \Psi_{\Gamma,f}[A]=\lim_{\Delta s \to 0}\,\Big(\sum_I \widehat{L}_I \,\Psi_{\Gamma,f}[A]\Big)\;.
	\label{eq:L on cyl}
\end{equation}
To be more specific, we need to define a consistent family of operators for finite $\Delta s$ and given cylindrical function. This step requires a number of choices which we state below. Then we can analyze the existence of the limit in the operator topology.

The construction of the operator $\widehat{L}_I$ for finite $\Delta s$ and given cylindrical function is discussed in section \ref{sec:elementary L}. It requires two ingredients: the \emph{two-hand} operator $\hat{Y}_{I\alpha\beta}^i$ corresponding to the quantization of the quantity introduced in (\ref{eq:regularized Y}), and the inverse of the local volume operator $\hat{V}_I$. These two ingredients enter in the definition of $\widehat{L}_I$ in the way specified by equations (\ref{eq:regularized G}) and (\ref{eq:regularized L_I}). In section \ref{sec:elementary L} we show that the operator $\widehat{L}_I$ can be implemented in such a way that, for given cylindrical function, the limit $\Delta s \to 0$ is well-defined and has a number of desirable properties. This defines an `elementary' length operator in a fashion analogous to the one discussed in section \ref{sec:quantization V} for the `elementary' volume operator $\hat{V}_n$. The length operator for an extended curve $\gamma$ can be defined in terms of the `elementary' length operator as discussed in section \ref{sec:extended length}.

\subsubsection{Quantization: the two-hand operator} 
The quantity $\sum_{\alpha\beta} Y_{I\alpha\beta}^i$ introduced in (\ref{eq:regularized Y}) can be quantized following a procedure analogous to the one discussed in some detail for the volume in section \ref{sec:quantization V}. 

We recall that the subscripts ${}_{I\alpha\beta}$ in $Y_{I\alpha\beta}^i$ represent a cubic cell $R_I$, and two square portions $S^1_{I\alpha}$ and $S^2_{I\beta}$ belonging respectively to the two faces $S^1_I$ and $S^2_I$ that share a side of the cube. Let's consider a cylindrical function with graph $\Gamma$ and focus on a node of $\Gamma$. The partition in cells $R_I$ is refined so that each cell can contain at most one node (see the discussion in section \ref{sec:quantization V}). Let's assume that a node is contained in the cell $R_I$ and that two links $e_1$ and $e_2$ originating at the node intersect the surfaces $S_{I\bar{\alpha}}^1$ and $S_{I\bar{\beta}}^2$ respectively. The flux operator for a surface $S$ acts non-trivially on a cylindrical function only if the surface $S$ is punctured by a link of $\Gamma$. As a result to the sum $\sum_{\alpha\beta} \hat{Y}_{I\alpha\beta}^i$ only the term $\hat{Y}_{I\bar{\alpha}\bar{\beta}}^i$ contributes. Let's focus on this term. At this point an adaptation of the curves involved in the definition of the quantity $V_{x_I}^{ijk}(\sigma,\sigma')$ defined in (\ref{eq:Vijk}) to the graph $\Gamma$ is invoked:
\begin{itemize}
	\item[-] the point $x_I$ is chosen to coincide with the position of the node,
	\item[-] the two curves $\gamma^1_{x_I\sigma}$ and $\gamma^2_{x_I\sigma}$ are adapted to the portions of the two links $e_1$ and $e_2$ contained in the cell $R_I$.
\end{itemize}
As a result the operator $\hat{Y}_{I\bar{\alpha}\bar{\beta}}^i$ has the diagrammatic representation shown in figure \ref{fig: regularization L}-(c). This adaptation to the graph $\Gamma$ has the remarkable property that the operator is independent of the size $\Delta s$ of the cell, thanks to the shrinking property discussed for the volume. As a result the limit $\Delta s\to 0$ can be taken trivially. 

The construction described above defines a two-hand operator on $\mc{K}_0(\Gamma)$ corresponding to the regularized quantity $\sum_{\alpha\beta} Y_{I\alpha\beta}^i$. The appropriate labeling for this operator is a node $n$ of $\Gamma$ and two links $e_1$, $e_2$ originating at the node. We call the triple $\{n,e_1,e_2\}$ a \emph{wedge}\footnote{The notion of wedge was introduced by Reisenberger in a different setting. The geometrical role played by the wedge here (see in particular section \ref{sec:extended length}) is close to the one of \cite{Reisenberger:1996ib}, however the use we will make of it is different.} of $\Gamma$ and indicate it with the symbol $w$. In the following we will refer to this operator as the two-hand operator $\hat{Y}^i(\gamma_w)$ defined as 
\begin{equation}
	\hat{Y}^i(\gamma_w)=\eps^{i\,j'k'}\;  \mc{D}^{(1)}(h_{e_1}[A])_{j'}^{\phantom{j'}j}\; \mc{D}^{(1)}(h_{e_2}[A])_{k'}^{\phantom{k'}k}\;\hat{F}_j(S_{e_1})\hat{F}_k(S_{e_2})
\end{equation}
It acts on a spin~network state belonging to $\mc{K}_0(\Gamma)$ in the following way:
\begin{align}
	&\hat{Y}^i(\gamma_w)\;\Psi_{\Gamma,j,i_k}[A]\,=\,
		\hat{Y}^i(\gamma_w)\, \Big(\mc{D}^{(j_1)}(h_{e_1}[A])_{m_1'}^{\ph{m_1}m_1}\cdot\cdot\, \mc{D}^{(j_L)}(h_{e_L}[A])_{m_L'}^{\ph{m_L}m_L}\;{v_k}^{(j_1\cdot\cdot j_L)}_{m_1\cdot\cdot m_L}\Big)\,\times \textrm{rest}^{m'_1\cdot\cdot m'_L} \label{eq:Yi on intertwiner}\\
&=(8\pi \gamma L_P^2)^2\, \mc{D}^{(j_1)}(h_{e_1}[A])_{m_1'}^{\ph{m_1}m_1}\cdot\cdot\, \mc{D}^{(j_L)}(h_{e_L}[A])_{m_L'}^{\ph{m_L}m_L}\;\Big(\eps^{ijk} T^{(j_1)\;\,m_1''}_{j\;m_1} T^{(j_2)\;\,m_2''}_{k\;m_2} \,{v_{k}}^{(j_1\,j_2\cdot\cdot j_L)}_{m_1''m_2''\cdot\cdot m_L}\Big)\,\times \textrm{rest}^{m'_1\cdot\cdot m'_L} \nonumber	
\end{align}
Notice that the construction of the operator $\hat{Y}^i(\gamma_w)$ is a crucial step in addressing difficulty (a). The key observation is that we can define a two-hand operator for each couple of links originating at a spin~network node. Choosing a couple of links identifies two surfaces dual to the links. Such two surfaces intersect at a curve which we have called $\gamma_w$. This is the way the operator knows about the curve. An equivalent way of identifying the curve is the following: a \emph{wedge}, i.e. a couple of links originating at a node, identifies a surface bounded by the two links. The dual to this surface is the curve we are looking for.

\subsubsection{Quantization: the inverse volume operator}\label{sec:inverse volume}
Now we address difficulty (b). The denominator in the regularized expression (\ref{eq:regularized G}) has been discussed in some detail in section \ref{sec:quantization V}. It has the meaning of volume of the cell $R_I$. At the quantum level it corresponds to the operator $\widehat{V}_I$ which is known to have a non-trivial kernel. The kernel consists of 
\begin{itemize}
	\item[-] spin~network states with graph $\Gamma$ which have no node contained in the region $R_I$,
	\item[-] specific superpositions over intertwiners of spin~network states with graph $\Gamma$ having a node in the region $R_I$.
\end{itemize}
The former class forms a huge space which has a clear geometrical interpretation in terms of the dual picture of quantum geometry described in section \ref{sec:picture}: nodes of the spin~network correspond to chunks of space of definite volume. No node no volume. As far as the latter class of states belonging to the kernel is concerned we have that $3$-valent nodes always correspond to zero volume \cite{Loll:1995tp}, while for higher valence a zero eigenvalue can be present or not depending on the spins on the incoming links. We call the zero eigenvalues of this latter class `accidental' zero eigenvalues.

Here we want to introduce an operator corresponding to the inverse of the volume of the cell $R_I$. Clearly the inverse of the operator $\widehat{V}_I$ does not exist and some generalization is needed. The simplest way out is to first restrict the domain of $\widehat{V}_I$ so that the resulting restricted operator is invertible, then to define an extension of its inverse to its maximal domain. Given the geometrical interpretation of the operator, our requirement for the operator corresponding to the inverse volume are stronger. We would like to satisfy the following two conditions:
\begin{itemize}
	\item preserve the dual picture of quantum geometry, i.e. the `inverse volume' operator acts \emph{at nodes}; as a result it has to annihilate spin~network states having no node contained in $R_I$,
	\item share the same eigenstates of $\widehat{V}_I$ and, for each non-vanishing eigenvalue of $\widehat{V}_I$, 
 have as eigenvalue the inverse of the corresponding one of $\widehat{V}_I$; this second requirement is expected to guarantee the appropriate semiclassical behaviour.
\end{itemize} 
These two requirements do not completely fix the matrix elements of the operator we want to define. A freedom in the choice of the eigenvalue corresponding to the `accidental' zero eigenvalues of $\widehat{V}_I$ is left\footnote{Classically the volume of a region can be arbitrarily small. As a result, the inverse volume can be arbitrarily big and the inverse can be extended in zero by continuity to be $+\infty$. On the other hand, choosing the eigenvalue corresponding to an `accidental' zero eigenvalue of $\widehat{V}_I$ by continuity is not an option as the spectrum of the volume operator is discrete. From this point of view any real number in the interval $[0,\infty[$ is equally good.\label{footnote:accidentalfreedom}}. In the following a specific choice is made.

\begin{figure}
\centering
\includegraphics[width=0.4\textwidth]{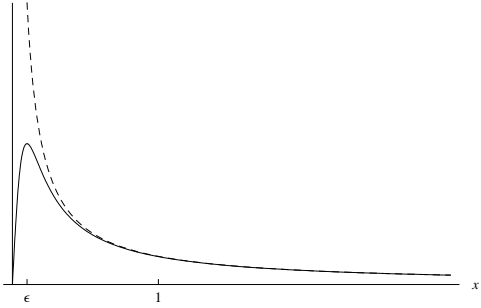}
\caption{Tikhonov regularization of the inverse. The plot shows the function $\frac{x}{x^2+\eps^2}$ (full line) and the function $x^{-1}$ defined for $x>0$ (dashed line). The limit $\eps\to 0$ defines an extension of the inverse of $x$ to the domain $x\geq 0$. Such extension vanishes in $x=0$ and coincides with $x^{-1}$ for $x>0$. This same property is shared by the eigenvalues of the operators $\widehat{V}$ and $\widehat{V^{-1}}$, as can be shown using spectral decomposition. 
}
\label{fig:Tikhonov}
\end{figure}

An operator which meets both the requirements described above can be easily introduced thanks to Tikhonov regularization \cite{Tikhonov:1943}. Let's notice that, for finite $\eps$, the inverse of the operator $\widehat{V}^2+\eps^2 L_P^6$ exists even if the inverse of $\widehat{V}^2$ does not exist. Then we define the operator $\widehat{V^{-1}}$ as the limit
\begin{equation}
	\widehat{V^{-1}}=\lim_{\eps\to 0}\, (\,\widehat{V}^2+\eps^2 L_P^6\, )^{-1} \, \widehat{V} \,.
	\label{eq:Tikhonov}
\end{equation}
Such limit exists and defines a hermitian operator $\widehat{V^{-1}}$ which commutes with $\widehat{V}$ and admits a self-adjoint extension to the whole Hilbert space $\mc{K}_0$. Such operator clearly annihilates spin~network states having no node in the region $R_I$ as $\widehat{V}$ does. Actually it has the same kernel of $\widehat{V}$ and the operator $1 - \widehat{V^{-1}}\widehat{V}$ is the projector on the kernel. Moreover, the non-vanishing eigenvalues of $\widehat{V^{-1}}$ are simply the inverse of the corresponding eigenvalues of $\widehat{V}$ as can be shown by spectral decomposition (see Fig. \ref{fig:Tikhonov}).

While the two requirements for the generalized inverse described above have a physical interpretation, the choice of using the Tikhonov regularization to define such an extension is to be considered only as a technical tool. Other choices are possible\footnote{See for instance footnote (\ref{footnote:accidentalfreedom}). On the other hand a weakened form of the second of the two requirements can be considered. For instance we could assume that the eigenvalues of the new operator coincide with the inverse of the eigenvalues of the volume only asymptotically. As an example, instead of taking the limit $\eps\to 0$ in the regularization discussed in the text, we could define an operator corresponding to $\eps=1$.  Mathematically such an operator would certainly not be called a `generalized' inverse. However we should recall that our purpose here is not to define \emph{the} `inverse volume' operator but to construct the length operator. An ingredient built in this way probably would not destroy the semiclassical behaviour of the length. From this point of view our choice of using Tikhonov regularization is to be considered as the minimal and simplest one.}. It is remarkable from this point of view that the extension $\widehat{V^{-1}}$ obtained through Tikhonov regularization of the inverse of $\widehat{V}$ satisfies the defining properties\footnote{The operator $\widehat{V^{-1}}$ defined through Tikhonov regularization of the inverse of $\widehat{V}$ satisfies the following properties: 
\begin{equation*}
\widehat{V}\, \widehat{V^{-1}}\, \widehat{V}=\widehat{V}\quad,\quad 
\widehat{V^{-1}}\, \widehat{V}\, \widehat{V^{-1}}=\widehat{V^{-1}}\quad,\quad (\widehat{V^{-1}}\, \widehat{V})^\dagger=\widehat{V^{-1}}\, \widehat{V}\quad,\quad (\widehat{V}\, \widehat{V^{-1}})^\dagger=\widehat{V}\, \widehat{V^{-1}}\;.
\end{equation*}
} of the Moore-Penrose pseudoinverse \cite{MoorePenrose:inv} 
 which is known to exist and to be unique.

The ingredient we need in order to build the `elementary' length operator is the local inverse volume $\widehat{V_I^{-1}}$ for the cell $R_I$. In this section we have shown that such operator can be introduced and that it acts non-trivially only at a node of the spin~network state. As a result we can define an inverse volume operator for a region $R_n$ dual to the node $n$. In the following we call it $\widehat{V^{-1}}(R_n)$.

\subsubsection{Quantization: the `elementary' length operator}\label{sec:elementary L}
The final step of the construction is to build an operator corresponding to the quantity $L_I$ defined by expression (\ref{eq:regularized L_I}) out of the two-hand operator $\sum_{\alpha \beta} \hat{Y}^i_{I\alpha \beta}$ and the local inverse-volume operator $\widehat{V_I^{-1}}$ as prescribed by equation (\ref{eq:regularized G}). Both of them, the two-hand operator and the local inverse-volume operator, admit a dual description in terms of nodes and links of the graph of a spin~network state. Such dual description matches with the desired one described in section \ref{sec:picture} for the length operator. Therefore in the following we will make use of it and call the operator built in such manner the `elementary' length operator. The length operator for an extended curve corresponding to the classical quantity (\ref{eq:regularized L}) will be discussed in section \ref{sec:extended length}.

Let's consider the Hilbert space $\mc{K}_0(\Gamma)$ spanned by spin~network states with graph $\Gamma$ and focus on a node $n$ of $\Gamma$ and two links $e$ and $e'$ originating at the node $n$. We call the triple $w=\{n,e,e'\}$ a \emph{wedge} of the graph $\Gamma$. To each wedge of $\Gamma$ we associate a curve $\gamma_w$ in the following way: two links $e$ and $e'$ sharing an endpoint have as dual two surfaces $S_e$, $S_{e'}$ which intersect at the curve $\gamma_w$. We say that the curve $\gamma_w$ is dual to the wedge $w$. An alternative way of identifying the curve $\gamma_w$ is to consider the wedge $w=\{n,e,e'\}$ as a soapfilm-like surface with boundary wire $e$, $e'$ and declare that the curve $\gamma_w$ is dual to the \emph{soap film} $w$.

The idea is to introduce on $\mc{K}_0(\Gamma)$ an operator which measures the length of the curve $\gamma_w$. As a result we have a length operator $\widehat{L}(\gamma_w)$ for each wedge $w=\{n,e,e'\}$ of the graph $\Gamma$. Such length operator is built out of the two-hand operator $\hat{Y}^i(\gamma_w)$ for the wedge $w$ and the local inverse-volume operator $\widehat{V^{-1}}(R_n)$ for the node $n$ belonging to the wedge $w$. The two operators enter in the definition of $\widehat{L}(\gamma_w)$ in a way corresponding to the classical expressions (\ref{eq:regularized L_I}) and (\ref{eq:regularized G}). As the two operators do not commute, a specific ordering has to be chosen. This has to be done so that the quantity corresponding to the argument of the square root in (\ref{eq:regularized L_I}) is a positive semi-definite hermitian operator. The ordering we study here is the following 
\begin{equation}
	\widehat{L(\gamma_{w})}=\frac{1}{2}\sqrt{\widehat{V^{-1}}(R_n)\; \delta_{ij}\, \widehat{Y}^i(\gamma_{w}) \widehat{Y}^j(\gamma_{w})\; \widehat{V^{-1}}(R_n)}\,.
	\label{eq:L hat1}
\end{equation}
Let's notice that if we introduce the non-hermitian operator $\hat{G}(\gamma_{w^i})$
\begin{equation}
\widehat{G}^i(\gamma_w)=\frac{1}{2}\,\hat{Y}^i(\gamma_w)\,\widehat{V^{-1}}(R_n)	\,,
	\label{eq:G hat}
\end{equation}
then the operator $\widehat{L}(\gamma_w)$ introduced above can be written as
\begin{equation}
	\widehat{L}(\gamma_w)=\sqrt{\delta_{ij}\, \widehat{G}^{i\dagger}(\gamma_w)\, \widehat{G}^j(\gamma_w)}\,.
	\label{eq:L hat2}
\end{equation}
This last expression shows that the argument of the square root is positive semi-definite and hermitian. As a result the operator $\widehat{L}(\gamma_w)$ is positive semi-definite and hermitian.

Other choices of ordering are possible. The choice discussed above has the technical advantage of being very simple to evaluate. A detailed analysis of how it acts at wedges of spin~network states having four-valent nodes will be presented in next section. 

Here we briefly recall that an ordering ambiguity is always present for composite operators in quantum mechanics \cite{Weyl:bookGroupsQM}. In the present case the requirements of hermitianity and positivity are not sufficient for completely fixing the ambiguity. Some extra physical input is needed. Different orderings are expected to modify the eigenstates and the eigenvalues in the deep quantum regime (namely, for eigenvalues of the order $L_P$). Ultimately it is only identifying a specific observable and probing experimentally the deep quantum regime that a specific ordering can be singled out. On the other hand different orderings are generally  not expected to change the semiclassical behaviour of the operator. In the following we mention two other ordering choices which we find interesting. 

Thanks to the fact that fluxes are generators of $SU(2)$ transformations, a Weyl ordering \cite{Weyl:bookGroupsQM} of the length operator in terms of fluxes could be viable\footnote{In \cite{Weyl:bookGroupsQM} Weyl suggested a way to resolve the ordering ambiguity in a group theoretical way. The idea is best expressed with a simple example: let's consider a particle of spin $J^i$ with $i=1,2,3$ and a composite classical observable $O(J^i)$. At the quantum level the spins do not commute $[\hat{J}^i,\hat{J}^j]=\hbar \eps_k^{\ph{k}ij}\hat{J}^k$ and an ordering ambiguity is present in the definition of the operator $\hat{O}$. This ambiguity can be resolved using the finite $SU(2)$ transformations $\hat{U}(\alpha_i)=e^{i \alpha_i \hat{J}^i}$. The strategy is the following. We consider the Fourier transform $\tilde{O}(\alpha_i)$ of $O(J^i)$ at the classical level, $\tilde{O}(\alpha_i)=\int d J^i O(J^i) e^{-i \alpha_i J^i} $. Then the operator $\hat{O}$ is defined as
\begin{equation*}
	\hat{O}=\int d\mu(\alpha_i) \,\tilde{O}(\alpha_i)\,\hat{U}(\alpha_i)\, , 
\end{equation*}
where $d\mu(\alpha_i)$ is the Haar measure for $SU(2)$. This prescription for the construction of the composite operator $\hat{O}$ is generally called \emph{Weyl ordering} as it amounts to a specific ordering of the operators $\hat{J}^i$. When viable, this construction is particularly satisfying because it automatically guaranties many desirable properties for composite operators. 
}

A different possibility is to regard the length operator as a composite operator made out of the hermitian \emph{triad} operator $\hat{e}^i(\gamma_w)$ so that $\widehat{L}(\gamma_w)=\sqrt{\delta_{ij} \hat{e}^i(\gamma_w) \hat{e}^j(\gamma_w)}$. Then the triad operator for a curve associated to the wedge $w$ can be defined in terms of the operator $\widehat{G}^i(\gamma_w)$ introduced in (\ref{eq:G hat}) as
\begin{equation}
\hat{e}^i(\gamma_w)=\frac{1}{2} \big(\widehat{G}^i(\gamma_w)+\widehat{G}^{i\dagger}(\gamma_w)\big)\,.
	\label{eq:e hat}
\end{equation}
This possibility is particularly appealing as it rests on the physical statement that the triad is a well-defined operator and measuring the length amounts to measuring a certain function of the triad. Moreover the triad also appears elsewhere, most notably in the Hamiltonian constraint \cite{Thiemann:1996aw,Thiemann:1996av,Thiemann:1997rv}.

The `elementary' length operator introduced above measures the length of a curve which is defined in terms of the graph $\Gamma$: it is the dual to the surface spanned by two links originating at a node. The operator for the length of an extended curve $\gamma$ can be written in terms of the `elementary' length operator. We will discuss in detail its construction in section \ref{sec:extended length}

\vspace{2em}

Before moving to the analysis of the properties of the `elementary' length operator, we briefly describe a different construction due to Thiemann \cite{Thiemann:1996at}. 

\subsection{Thiemann's construction}\label{sec:thiemann}
In a series of remarkable papers \cite{Thiemann:1996aw,Thiemann:1996av,Thiemann:1997rv}, Thiemann proposed a technique for quantizing the quantity $G_i(s)$, namely the pull-back of the triad along a curve $\gamma$, in terms of the holonomy and the volume operator. The technique rests on the following relation holding at the classical level
\begin{equation}
	G_i(s)= \frac{2}{8\pi \gamma G_N/c^3} \{A_a^i(\gamma(s))\dot{\gamma}^a(s), V(R)\}
	\label{eq:G as poisson bracket}
\end{equation}
where $V(R)$ is the volume of a region containing the point $x^a=\gamma^a(s)$. This classical expression suggests a remarkably simple way to promote $G_i(s)$ to an operator acting on a cylindrical function with given graph $\Gamma$. One starts \emph{holonomizing}\footnote{That is writing a component of the connection in a point as the limit $\Delta s \to 0$ of a function of the holonomy along a curve $\gamma^a_I(s)$ with $s\in [0, \Delta s]$.} it, and then promoting the holonomy and the volume to operators and the Poisson bracket to a commutator: 
\begin{equation}
	\hat{G}_i(s_I)\tau^i=\frac{2}{8\pi \gamma L_P^2}\,\lim_{\Delta s \to 0}\frac{1}{\Delta s}\big(h_{\gamma_I}^{-1}\hat{V}_I h_{\gamma_I} - \hat{V}_I \big)\;.
	\label{eq:Thiemann trick}
\end{equation}
Notice that no inverse of the volume operator is needed as it appears only linearly. At this point an adaptation of the curve $\gamma_I$ to a link of the graph $\Gamma$ is invoked. No fluxization procedure is required as the field $E^a_i(x)$ is contained only in the expression of the volume which has already been defined\footnote{Notice that the Ashtekar-Lewandowski version of the volume operator is used \cite{Ashtekar:1997fb}. See also \cite{Giesel:2005bm,Giesel:2005bk} for a set of consistency check.}. When using this technique for constructing an `elementary' length operator \cite{Thiemann:1996at} one ends up with an object which acts at nodes of the spin~network state and is supposed to measure a length in the direction of a link. 

The asset of using this technique to introduce a length operator is that it deals with both difficulty (a) and difficulty (b) at one stroke. However, one ends up with an operator acting in a way which fits very badly in the dual picture of quantum geometry described in section 1. This fosters the suspicion that this operator actually measures something with dimensions of length which could be unrelated to the length of a curve. An analysis of its eigenstates and its spectrum would clarify this issue and discriminate if its proper quantum geometrical meaning is that of length or not. Unfortunately, this issue has not been sufficiently explored and the answer is unclear. 

\section{Properties of the `elementary' length operator}\label{sec:properties}
The `elementary' length operator measures the length of a curve defined in the following intrinsic way. We start with a spin~network state with graph $\Gamma$ and focus on a node and a couple of links originating at that node. Then we consider two surfaces dual to the two links. These two surfaces intersect at a curve. The `elementary' length operator measures the length of this curve. 

More formally, we consider the Hilbert space $\mc{K}_0(\Gamma)$ and - for each wedge $w=\{n,e,e'\}$ of the graph $\Gamma$ - we define an operator $\widehat{L}(\gamma_w)$ which measures the length of a curve $\gamma_w$ associated to the wedge $w$. Clearly, there is a whole class of curves which are dual to the wedge $w$. The `elementary' length operator has the property of depending only on such class.

  The operator $\widehat{L}(\gamma_w)$ goes from $\mc{K}_0(\Gamma)$ to itself, that is it does not change the graph $\Gamma$. It acts on the $L$ links $e_1\mdots e_L$ originating at the node $n$ distinguishing the links $e,e'$ from the remaining $L-2$ links. Moreover, when acting on the spin~network basis, it does not change the spins associated to the links. As a result it can act non-trivially only on the intertwiner associated to the node $n$. In formulae, we have that for the wedge $w_{12}=\{n,e_1,e_2\}$
\begin{align}
	\widehat{L}(\gamma_{w_{12}})\;\Psi_{\Gamma,j,i_k}[A]\,=&\,
		\widehat{L}(\gamma_{w_{12}})\, \Big(\mc{D}^{(j_1)}(h_{e_1}[A])_{m_1'}^{\ph{m_1}m_1}\cdot\cdot\, \mc{D}^{(j_L)}(h_{e_L}[A])_{m_L'}^{\ph{m_L}m_L}\;{v_k}^{(j_1\cdot\cdot j_L)}_{m_1\cdot\cdot m_L}\Big)\,\times \textrm{rest}^{m'_1\cdot\cdot m'_L} \nonumber\\
&\hspace{-3em}=\mc{D}^{(j_1)}(h_{e_1}[A])_{m_1'}^{\ph{m_1}m_1}\cdot\cdot\, \mc{D}^{(j_L)}(h_{e_L}[A])_{m_L'}^{\ph{m_L}m_L}\;\Big(\sum_h(L_{w_{12}})_k^{\ph{k}h}\,{v_{h}}^{(j_1\cdot\cdot j_L)}_{m_1\cdot\cdot m_L}\Big)\,\times \textrm{rest}^{m'_1\cdot\cdot m'_L}\nonumber\\
&\hspace{-3em}=\sum_h (L_{w_{12}})_k^{\ph{k}h} \;\Psi_{\Gamma,j,i_h}[A]  	\label{eq:L on intertwiner}
\end{align}
that is the action of the operator is completely encoded into the matrix $(L_{w_{12}})_k^{\ph{k}h}$.

In this section we compute the matrix elements of this operator for nodes which are four-valent and discuss some of its properties. In the particular we compute a number of its eigenvalues and eigenstates, discuss some non-trivial commutation relations and analyse its semiclassical behaviour.\\

Let's start recalling some basics about the intertwiner vector space (see for instance \cite{Landau:1960} ch. XIV). We consider the vector space $\mc{V}^{(j)}$ where the representation $j$ of $SU(2)$ acts. The basis $|j,m\rangle$ with $m=-j\mdots +j$ is such that the generators $\hat{T}_i$ of $SU(2)$ have matrix elements $\hat{T}_i|j,m\rangle=\sum_{m'} T^{(j)m'}_{i\,m}|j,m'\rangle$ and $|j,m\rangle$ is a basis of eigenstates of $\hat{T}_0$ to the eigenvalue $m$ and of the Casimir operator $\de^{ij}\hat{T}_i\hat{T}_j$ to the eigenvalue $j(j+1)$. 
Then we consider the tensor product $\mc{V}^{(j_1\ldots j_L)}= \mc{V}^{(j_1)} \otimes \cdots \otimes \mc{V}^{(j_L)}$ of $L$ irreducible representations of $SU(2)$ with spins $j_1\mdots j_L$. This is a vector space of dimension $(2j_1+1)\times\cdots\times (2j_L+1)$ which can be decomposed into a sum of irreducible components. Here we are interested in the subspace $\mc{V}_0^{(j_1..j_L)}$ that transforms in the trivial representation. This space is $K$ dimensional, where K is the multiplicity with which the trivial representation appears in the decomposition. \emph{Intertwiners} form a basis of this vector space. In terms of the natural basis $|j_1,m_1\rangle\cdot\cdot\,|j_L,m_L\rangle$ of $\mc{V}^{(j_1\ldots j_L)}$, the intertwiner basis can be written in the following way
\begin{equation}
|k\,\rangle=\sum_{m_1..m_L}{v_k}^{(j_1\ldots j_L)}_{m_1\ldots m_L}\;\, |j_1,m_1\rangle\cdot\cdot\,|j_L,m_L\rangle\;.
	\label{eq:intertwiner}
\end{equation}
We call the coefficients ${v_k}^{(j_1\ldots j_L)}_{m_1\ldots m_L}$ intertwining tensor. The intertwiner vector space inherits a scalar product from the one defined on $\mc{V}^{(j)}$, namely $\langle j,m|j,m'\rangle=\delta_{m,m'}$. With respect to this scalar product, the states $|k\,\rangle$ with $k=1\mdots K$ are chosen so that $\langle h|k\rangle=\delta_{k,h}$. Therefore they form an orthonormal basis of $\mc{V}_0^{(j_1..j_L)}$. At the level of intertwining tensors this amounts to the following relation
\begin{equation}
	\eta^{(j_1)\, m'_1 m_1}\cdot\cdot\, \eta^{(j_L)\, m'_L m_L}\;{v_h}^{(j_1\ldots j_L)}_{m'_1\ldots m'_L}\;{v_k}^{(j_1\ldots j_L)}_{m_1\ldots m_L}=\delta_{k,h}
	\label{eq:intertwiners scalar product}
\end{equation}
where $\eta^{(j)\, m n}$ is the Wigner metric tensor and is given by $\eta^{(j)\,mn}=(-1)^{j-m}\delta_{m,-n}$. A summation over repeated $m$ indices is understood. In the trivalent case $L=3$, for admissible\footnote{\label{foot:admissible}The spins $j_1, j_2, j_3$ are said to be admissible if $j_1+j_2+j_3$ is integer and they satisfy the triangular inequality $|j_1-j_2|\leq j_3 \leq j_1+j_2$. These two conditions are equivalent to the requirement that there are three integers $a,b,c$ such that $j_1=(a+b)/2$, $j_2=(a+c)/2$, $j_3=(b+c)/2$. This parametrization often turns out to be useful.} spins $j_1, j_2, j_3$, the intertwiner vector space is one-dimensional and the unique intertwining tensor is given by the Wigner $3j$ symbol
\begin{equation}
	v^{(j_1\,j_2\,j_3)}_{m_1\,m_2\,m_3}=
	\left(\begin{array}{ccc}
	j_1 & j_2 & j_3\\
	m_1 & m_2 & m_3
	\end{array}\right)\;.
	\label{eq:3j}
\end{equation}
The expression of the Wigner $(3j)$ symbol can be found for instance in \cite{Landau:1960}. The intertwining tensor vanishes if the three spins are not admissible. In the four-valent case, the intertwiner vector space is less trivial. An orthonormal basis can be given in terms of the coupling of two three-valent intertwiners
\begin{equation}
	{v_k}^{(j_1\,j_2)\,(j_3\,j_4)}_{m_1\,m_2\,m_3\,m_4}=\sqrt{2k+1}\; \eta^{(k)n n'}\, v^{(j_1\,j_2\,k)}_{m_1\,m_2\,n}\;v^{(k\,j_3\,j_4)}_{n'\,m_3\,m_4}
	\label{eq:four valent tensor}
\end{equation}
The basis is labeled by the spin $k$ of the coupling channel. The dimension of the vector space is given by the number of the admissible spins $k$. As is well known, orthonormal basis associated to different coupling channels are related by a unitary transformation given by the Racah $\{6j\}$ symbol \cite{Landau:1960}. When using the bra-ket notation we will indicate the coupling channel explicitly. For instance the basis state associated to the intertwining tensor given in equation (\ref{eq:four valent tensor}) will be denoted $|k_{12}\,\rangle$.

A planar diagrammatic notation often turns out to be useful. Here we follow \cite{Varsh:1988}. We represent the Wigner metric  and  the  three-valent intertwining tensor by an oriented line and by a node with three links oriented counter-clockwise\footnote{A minus sign in place of the $+$ will be used to indicate clockwise orientation of the links.}, respectively:
\begin{equation}
	\eta^{(j)m n}=\; \parbox[t]{80pt}{\includegraphics{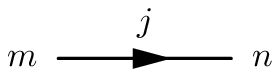}}\;\;,  \hspace{4em} v^{(j_1\,j_2\,j_3)}_{m_1\,m_2\,m_3}=\; \parbox[c]{100pt}{\includegraphics{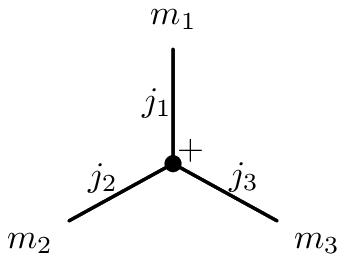}}\;\;.
	\label{eq:eta and v diagram}
\end{equation}
Moreover we denote the matrix elements $T^{(j)m'}_{i\,m}$ of the $SU(2)$ generators by an oriented line in representation $j$ with a grasping, i.e with a cross and an outcoming line in representation $1$  
\begin{equation}
	T^{(j)m'}_{i\,m}\equiv \;\; \parbox[b]{100pt}{\includegraphics{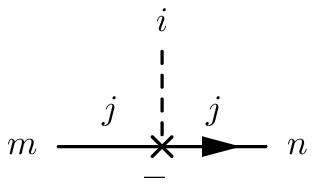}\\[-1.8em]} \!\!=\;\; i N_{(j)} \;\; \parbox[b]{100pt}{\includegraphics{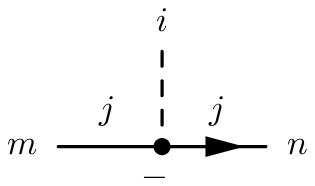}\\[-1.8em]}. 
	\label{eq:T diagram}
\end{equation}
In the last equality of (\ref{eq:T diagram}) we have highlighted the fact that the grasping can be expanded on (the unique state of) the intertwiner basis of $\mc{V}^{(j\,1\,j)}$. 
We have that $N_{(j)}=\sqrt{j(j+1)(2j+1)}$. In the following we will always denote a link in the adjoint representation by a dashed line and will suppress the label `$1$'.

Adopting this diagrammatic notation, the state $|k_{12}\rangle$ associated to the intertwining tensor (\ref{eq:four valent tensor}) is given by
\begin{equation}
	|k_{12}\,\rangle=\sqrt{2k+1} \parbox[c]{100pt}{\includegraphics{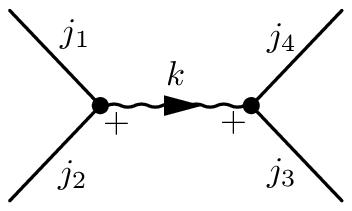}}
	\label{eq:k12}
\end{equation}
where a wiggly line has been used to denote the virtual link associated to the coupling channel. The states $|k_{12}\,\rangle$ with\footnote{We have that $k_{\textrm{min}}=\textrm{max}(|j_1-j_2|,|j_3-j_4|)$ and $k_{\textrm{max}}=\textrm{min}(j_1+j_2, j_3+j_4)$.} $k=k_{\textrm{min}},k_{\textrm{min}}+1\mdots k_{\textrm{max}}-1, k_{\textrm{max}}$ form an orthonormal basis of the four-valent intertwiner vector space $\mc{V}_0^{(j_1..j_4)}$. 

In the following section we consider a graph $\Gamma$ containing a four-valent node $n$ which is source for the links $e_1\mdots e_4$ and derive the matrix elements of the `elementary' length operator $\widehat{L}(\gamma_{w_{12}})$ on the spin~network basis. We choose the spin~network basis $\Psi_{\Gamma,j,i}$ of $\mc{K}_0(\Gamma)$ so that the node $n$ is labeled by the intertwining tensor (\ref{eq:four valent tensor}). Thanks to (\ref{eq:L on intertwiner}) we have that the matrix elements of $\widehat{L}(\gamma_{w_{12}})$ on this basis coincide with the matrix elements of an operator $\hat{L}_{w_{12}}$ on the basis $|k_{12}\rangle$ of the intertwiner vector space $\mc{V}_0^{(j_1..j_4)}$,
\begin{equation}
	\big(\Psi_{\Gamma,j,i_h}, \widehat{L}(\gamma_{w_{12}}) \Psi_{\Gamma,j,i_{k}}\big)=\langle\, h_{12} |\hat{L}_{w_{12}}|k_{12}\rangle=(L_{w_{12}})_k^{\ph{k}h}
	\label{eq:( ,L )=<h,L k>}
\end{equation} 
The operator $\hat{L}_{w_{12}}$ defined in the four-valent intertwiner vector space is derived in the following section starting from the two-hand and the three-hand operator.

\subsection{Matrix elements of the length operator: four valent nodes}
The operator $\widehat{L}(\gamma_{w_{12}})$ is defined by equation (\ref{eq:L hat1}). In order to derive its matrix elements we need two ingredients, namely the matrix elements of the operator $\de_{ij}\, \hat{Y}^i(\gamma_w) \hat{Y}^j(\gamma_w)$ and the matrix elements of the inverse-volume operator $\widehat{V^{-1}}(R_n)$.

Let's introduce the operator $\hat{Y}^i_{w_{12}}: \mc{V}_0^{(j_1..j_4)}\to \mc{V}_0^{(1,j_1..j_4)}$ defined diagrammatically as follows
\begin{equation}
	\hat{Y}^i_{w_{12}} \parbox[c]{100pt}{\includegraphics{feyn.k.channel.eps}} = (8\pi \gamma L_P^2)^2\, \sqrt{3!} \;\;\parbox[c]{110pt}{\includegraphics{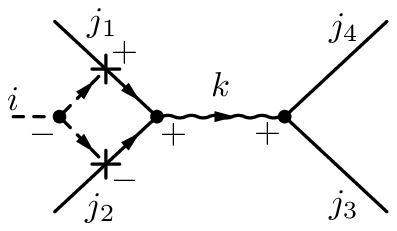}}\;\;.
	\label{eq:Y feyn}
\end{equation}
Then, using standard recoupling techniques, it's easy to show that the operator $\de_{ij}\hat{Y}^i_{w_{12}}\hat{Y}^j_{w_{12}}= (8\pi \gamma L_P^2)^4\, \hat{H}_{w_{12}}$ is defined from $\mc{V}_0^{(j_1..j_4)}$ to itself and has the following representation\footnote{The matrix elements $(H_{w_{12}})_k^{\ph{k}h}$ are given by the components of $\hat{H}_{w_{12}}|k_{12}\rangle$ on the basis $|k_{12}\rangle$. These can be computed evaluating the diagram

\vspace{-1.5em}

\begin{equation}
	\langle h_{12}|\hat{H}_{w_{12}}|k_{12}\rangle=3!\;\sqrt{2k+1}\sqrt{2h+1}\;\;\parbox[c]{110pt}{\includegraphics[scale=.8]{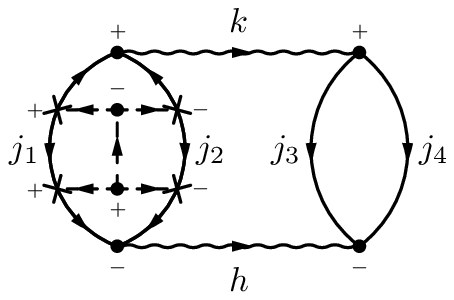}}\;\;.
	\label{eq:feyn H}
\end{equation}
The operator $\hat{H}_{w_{12}}$ turns out to be diagonalized by the basis $|k_{12}\rangle$.}
\begin{equation}
	\de_{ij}\hat{Y}^i_{w_{12}}\hat{Y}^j_{w_{12}}=(8\pi \gamma L_P^2)^4\,\sum_{k\,h} (H_{w_{12}})_k^{\ph{k}h} |h_{12}\rangle\langle k_{12}| =(8\pi \gamma L_P^2)^4\,\sum_k c(k,j_1,j_2) |k_{12}\rangle\langle k_{12}| 
	\label{eq:H12}
\end{equation}
with the coefficients $c(k,j_1,j_2)$ given by 
\begin{align}
c(k,j_1,j_2)=&\,j_1(j_1+1)\,j_2(j_2+1)-\left(\frac{k(k+1)-j_1(j_1+1)-j_2(j_2+1)}{2}\right)^2+\nonumber\\
 &-\frac{k(k+1)-j_1(j_1+1)-j_2(j_2+1)}{2}\;.	\label{eq:c coeff}
\end{align}
For admissible $k,j_1,j_2$ we have that $c(k,j_1,j_2)$ is non-negative as can be shown using the integer parametrization of footnote \ref{foot:admissible}. Therefore the operator $\hat{H}_{w_{12}}$ is hermitian and positive semi-definite.

Thanks to equation (\ref{eq:Yi on intertwiner}) we have that the matrix elements of the operator $\de_{ij}\widehat{Y}^i(\gamma_{w_{12}})\widehat{Y}^j(\gamma_{w_{12}})$ on the spin~network basis coincide with $(8\pi \gamma L_P^2)^4$ times the matrix elements of the operator $\hat{H}_{w_{12}}$ on the intertwiner basis.

A second ingredient we need is the operator $\hat{Q}_{\{e_1,e_2,e_3\}}$ from $\mc{V}_0^{(j_1..j_4)}$ to itself defined diagrammatically as follows
\begin{equation}
	\hat{Q}_{\{e_1,e_2,e_3\}} \parbox[c]{110pt}{\includegraphics{feyn.k.channel.eps}}= (8\pi \gamma L_P^2)^3\, \sqrt{3!} \parbox[c]{110pt}{\includegraphics{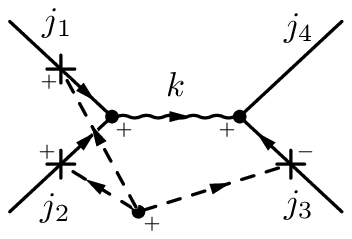}}
	\label{eq:Q feyn}
\end{equation}
Again, using standard recoupling techniques\footnote{The matrix elements $(Q_{\{e_1,e_2,e_3\}})_k^{\ph{k}h}$ are given by the components of $\hat{Q}_{\{e_1,e_2,e_3\}}|k_{12}\rangle$ on the basis $|k_{12}\rangle$. These can be computed evaluating the following diagram

\vspace{-1.8em}

\begin{equation}
	\langle h_{12}|\hat{Q}_{\{e_1,e_2,e_3\}}|k_{12}\rangle= (8\pi \gamma L_P^2)^3\,\sqrt{3!}\;\sqrt{2k+1}\sqrt{2h+1}\;\;\parbox[c]{110pt}{\includegraphics[scale=.8]{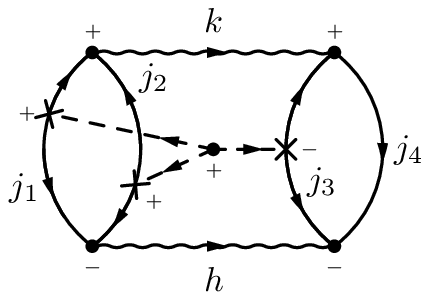}}\;\;.
	\label{eq:feyn Q}
\end{equation}
The matrix elements turn out to vanish unless $k=h\pm 1$. Using standard recoupling techniques the diagram in (\ref{eq:feyn Q}) can be written as the sum of three terms given by the product of four $\{6j\}$ symbols. As each $\{6j\}$ symbol contains either a spin $j=1$ opposite to a couple of equal spins or three spins $j=1$ in the same row, the explicit expression (\ref{eq:a(k,j)}) is available.} \cite{DePietri:1996pj} the matrix elements of $\hat{Q}_{\{e_1,e_2,e_3\}}$ can be found explicitly. On the basis $|k\rangle_{12}$ the operator has the following form
\begin{equation}
	\hat{Q}_{\{e_1,e_2,e_3\}}=(8\pi \gamma L_P^2)^3\,  \sum_{k=k_{\textrm{min}}+1}^{k_{\textrm{max}}} i\, a(k,j_1,j_2,j_3,j_4) \Big(|k_{12}\rangle\langle (k-1)_{12}|\,-\,|(k-1)_{12}\rangle \langle k_{12}|\Big)\;.
\end{equation}
The coefficients $a(k,j_1,j_2,j_3,j_4)$ are real and admit the following explicit expression\footnote{In the monochromatic case, i.e. for $j_1=j_2=j_3=j_4\,\equiv j_0$, a simpler expression for the coefficient $a$ is available
\begin{equation*}
	a(k,j_0)=\frac{k^2((2j_0+1)^2-k^2)}{4\sqrt{4k^2-1}}
\end{equation*}
with $k=0,1\mdots 2 j_0$.} \cite{Thiemann:1996au,Brunnemann:2004xi}
\begin{align}
a(k,j_1,j_2,j_3,j_4)=&\;\frac{1}{{4\sqrt{4k^2-1}}}\sqrt{(k+j_1+j_2+1)(-k+j_1+j_2+1)(k-j_1+j_2)(k+j_1-j_2)}\;\times	\nonumber\\
&\quad\times\;\sqrt{(k+j_3+j_4+1)(-k+j_3+j_4+1)(k-j_3+j_4)(k+j_3-j_4)}\;. \label{eq:a(k,j)}
\end{align}
The $\hat{Q}_{\{e_1,e_2,e_3\}}$ operator can be diagonalized
\begin{equation}
	\hat{Q}_{\{e_1,e_2,e_3\}}|q_i\rangle=(8\pi \gamma L_P^2)^3\,q_i \,|q_i\rangle
\end{equation}
and the eigenvalues $ (8\pi \gamma L_P^2)^3 \,q_i$ are real and non-degenerate. The eigenstates $|q_i\rangle$ with $i=1\mdots K$ provide an orthonormal basis of the intertwiner vector space. The non-zero eigenvalues always appear in pairs with opposite sign. A zero eigenvalue is present only when the dimension of the intertwiner vector space is odd \cite{Thiemann:1996au,Brunnemann:2004xi}. 

The operator $\hat{Q}_{\{e_1,e_2,e_3\}}$ is symmetric under even permutation of the labels $e_1,e_2,e_3$ and antisymmetric under odd permutations. Moreover the operators $\hat{Q}_{\{e_1,e_2,e_3\}}$, $\hat{Q}_{\{e_1,e_2,e_4\}}$, $\hat{Q}_{\{e_1,e_3,e_4\}}$, $\hat{Q}_{\{e_2,e_3,e_4\}}$ can be simultaneously diagonalized and actually coincide up to a sign. This is a straightforward consequence of the invariance of the four-valent intertwiner under $SU(2)$ transformations and the fact that the grasping operator is its generator. As a result the volume operator on the intertwiner vector space is simply given by
\begin{equation}
	\hat{V}_n=(8\pi \gamma L_P^2)^\frac{3}{2} \sum_i \sqrt{\frac{4}{8\times 3!}} \sqrt{|q_i|} \; |q_i\rangle \langle q_i|\;.
\end{equation}
The non-zero eigenvalues are twice degenerate. A non-degenerate zero eigenvalue is present when the dimension of the intertwiner vector space is odd. These are the `accidental' zeros of section \ref{sec:inverse volume}. In figure \ref{fig:volume length spectrum mono}-(a) some eigenvalues of $\hat{V}_n$ in the monochromatic case $j_1=j_2=j_3=j_4$ are reported. Now, let's define the operator $\hat{\Lambda}_n$ as
\begin{equation}
	\hat{\Lambda}_n={\sum_i}' \Big(\sqrt{\frac{4}{8\times 3!}} \sqrt{|q_i|}\Big)^{-1} \; |q_i\rangle \langle q_i|
\end{equation}
where the prime stands for sum over $i$ restricted to non-zero $q_i$. We have that the matrix elements on the spin~network basis of the inverse-volume operator $\widehat{V^{-1}}(R_n)$ discussed in section \ref{sec:inverse volume} coincide with $(8\pi \gamma L_P^2)^{-\frac{3}{2}}$ the matrix elements of $\hat{\Lambda}_n$ on the intertwiner basis.

Therefore the square of the `elementary' length operator $\widehat{L}(\gamma_{w_{12}})$ on the spin~network basis are given by $8\pi \gamma L_P^2$ times the matrix elements of the operator $\hat{\Lambda}_n \hat{H}_{w_{12}}\hat{\Lambda}_n$ on the intertwiner basis
\begin{equation}
	\big(\Psi_{\Gamma,j,i_h}, \big(\widehat{L}(\gamma_{w_{12}})\big)^2\, \Psi_{\Gamma,j,i_{k}}\big)=8\pi \gamma L_P^2 \langle\, h_{12} |\,\hat{\Lambda}_n \hat{H}_{w_{12}}\hat{\Lambda}_n\,|k_{12}\,\rangle\;.
\end{equation} 
The operator $\hat{\Lambda}_n \hat{H}_{w_{12}}\hat{\Lambda}_n$ is hermitian and positive semi-definite. As a result it can be diagonalized and its eigenvalues are positive. Lets call them $l_i^2$ with $i=1\mdots K$. Then we have 
\begin{equation}
	\hat{\Lambda}_n \hat{H}_{w_{12}}\hat{\Lambda}_n=\sum_i \,l_i^2\, |l_i\rangle \langle l_i|\;.
\end{equation}
The operator $\hat{L}_{w_{12}}$ is defined as $(8\pi \gamma L_P^2)^\frac{1}{2}$ times the square root of the operator $\hat{\Lambda}_n \hat{H}_{w_{12}}\hat{\Lambda}_n$, that is $\hat{L}_{w_{12}}=(8\pi \gamma L_P^2)^\frac{1}{2} \sum_i \,l_i\, |l_i\rangle \langle l_i|$. Therefore the matrix elements (\ref{eq:( ,L )=<h,L k>}) are given by 
\begin{equation}
	(L_{w_{12}})_k^{\ph{k}h}= (8\pi \gamma L_P^2)^\frac{1}{2} \sum_i \,l_i\, \langle h_{12}|l_i\rangle \langle l_i|k_{12}\rangle\;.
\end{equation}

\subsection{Spectrum and eigenstates: the four-valent monochromatic case}\label{sec:spectrum}
The `elementary' length operator $\widehat{L}(\gamma_{w_{12}})$ is diagonalized by spin~network states which have the node $n$ labeled by eigenstates of the operator $\hat{L}_{w_{12}}$
\begin{equation}
	\widehat{L}(\gamma_{w_{12}})\Psi_{\Gamma,j,i_{l_i}}=(8\pi \gamma L_P^2)^\frac{1}{2} \,l_i\; \Psi_{\Gamma,j,i_{l_i}}\;.
	\label{eq:L eigenvalue}
\end{equation}
In this brief section we report the eigenvalues of the `elementary' length operator in the case of a four-valent node with spins which are all equal, $j_1=j_2=j_3=j_4\equiv j_0$. The dimension of the four-valent \emph{monochromatic} intertwiner vector space $\mc{V}_0^{(j_0 j_0 j_0 j_0 )}$ is $K= 2j_0+1$.\\

\begin{figure}[t]
\centering
\begin{minipage}[b]{0.4\textwidth}
\begin{tabular}{c}
\includegraphics[height=.6\textwidth]{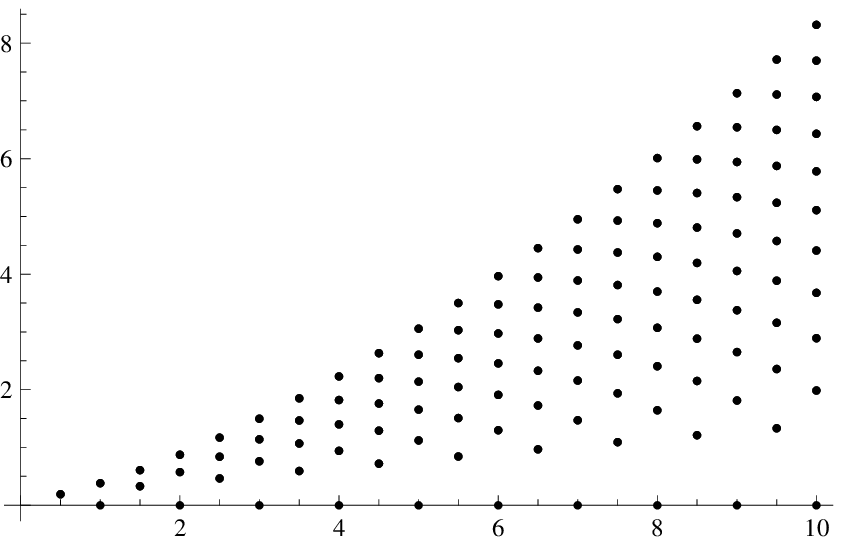}\\
(a)
\end{tabular}
\end{minipage}
\begin{minipage}[b]{0.1\textwidth}
\hspace{1em}
\end{minipage}
\begin{minipage}[b]{0.4\textwidth}
\begin{tabular}{c}
\includegraphics[height=.6\textwidth]{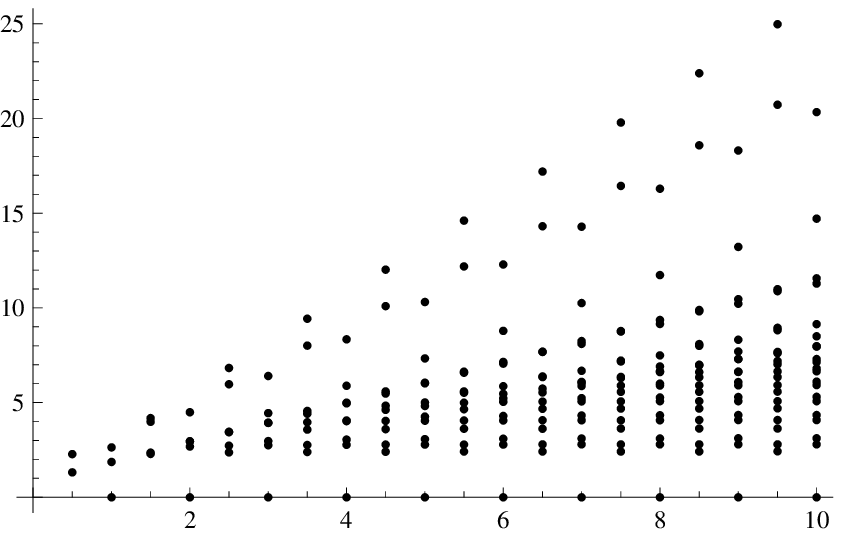}\\
(b)
\end{tabular}
\end{minipage}
\label{fig:volume length spectrum mono}
\caption{(a) Eigenvalues of the volume operator in units $(8\pi \gamma L_P^2)^{3/2}$. A monochromatic four-valent node has been considered. The $(2j_0+1)$ eigenvalues are plotted as a function of the half-integer spin $j_0$. Degeneracy to be taken into account. The maximum eigenvalue scales as $j_0^{3/2}$.  (b) Eigenvalues of the length operator for a wedge of monochromatic four-valent node. Units $(8\pi \gamma L_P^2)^{1/2}$ are used. The $(2j_0+1)$ eigenvalues are plotted as a function of the spin $j_0$. The maximum eigenvalue scales linearly in $j_0$.}
\end{figure}

\noindent For $j_0=\frac{1}{2}$ we have
\begin{align*}
l_1=3^{1/4} \hspace{3em} & |l_1\rangle=|0_{12}\rangle  \hspace{3em} \\
l_2=3^{3/4} \hspace{3em} & |l_2\rangle=|1_{12}\rangle
\end{align*}
For $j_0=1$ we have
\begin{align*}
l_1=0   \hspace{3em} & |l_1\rangle=\frac{\sqrt{5}}{3}\,|0_{12}\rangle+\frac{2}{3}\,|2_{12}\rangle\\
l_2=\sqrt{2}\times 3^{1/4} \hspace{3em}  & |l_2\rangle=-\frac{2}{3}\,|0_{12}\rangle+\frac{\sqrt{5}}{3}\,|2_{12}\rangle\\
l_2=2\times 3^{1/4} \hspace{3em}  & |l_2\rangle=|1_{12}\rangle
\end{align*}
For larger $j_0$ the eigenvalues and the eigenvectors can be easily found numerically. In figure \ref{fig:volume length spectrum mono}-(b) we report the series of eigenvalues as a function of $j_0$ for $j_0=\frac{1}{2},1,\frac{3}{2}\mdots 10$ . As expected by general arguments (see section \ref{sec:inverse volume}) a zero eigenvalue is present for $j_0$ integer. A preliminary investigation indicates that the eigenvalues are non-degenerate and that the value of the maximum eigenvalue scales linearly in $j_0$. In figure \ref{fig:volume length spectrum mono}-(a) we report also the analogous plot of the spectrum of the volume for comparison.

\subsection{Non-trivial commutators}\label{sec:commutators}
As explained before, when acting on a spin~network state, the length operator for the wedge $w=\{n,e,e'\}$ acts non-trivially only on the intertwiner space at the node $n$. As a result it commutes with the area operator for a surface dual to a link and with the volume operator $\hat{V}(R_{n'})$ for a region dual to a node $n'$ different from $n$. As far as the commutator with the volume operator $\hat{V}(R_n)$ is concerned, we have
\begin{equation}
	[\,\widehat{L}(\gamma_{\{n,e,e'\}})\,,\,\widehat{V}(R_n)\,]\neq 0\;.
	\label{eq:[L,V]}
\end{equation} 
The commutator can be computed, but its expression is not particularly enlightening. As an explicit check of non-triviality, we can evaluate the commutator in (\ref{eq:[L,V]}) on a spin~network state with a monochromatic four-valent node with low spin $j_0$. The problem reduces to a computation of the commutator of $(2 j_0+1)\times (2 j_0+1)$ matrices. For $j_0=\frac{1}{2}$ and $j_0=1$ we have that the commutator vanishes. The lowest-dimensional non-trivial case is for $j_0=\frac{3}{2}$.

Now we move on to commutators of elementary length operators. As the elementary length operator acts non-trivially only on the intertwiner space at a single node, we have that the commutator vanishes when we consider curves $\gamma_w$ and $\gamma_{w'}$ dual to wedges $w$ and $w'$ which do not contain the same node:
\begin{equation}
[\,\widehat{L}(\gamma_{\{n,e,e'\}})\,,\,\widehat{L}(\gamma_{\{\tilde{n},\tilde{e},\tilde{e}'\}})\,]= 0\qquad \textrm{if}\;\;n\neq \tilde{n}.
	\label{eq:[L,L]}
\end{equation} 
On the other hand, the commutator (\ref{eq:[L,L]}) is in general non-trivial if the two nodes coincide. In particular, for a four-valent node $n$ which is source for the links $e_1, e_2, e_3, e_4$, we have that there are six wedges $w_{12}, w_{13}, w_{14}, w_{23}, w_{24}, w_{34}$ and correspondingly six curves $\gamma_w$ dual to such wedges. As a check of non-triviality of the commutator we can look at its matrix elements in the monochromatic sector. In the previous sections we have studied the matrix elements of the operator $\hat{L}_{w_{12}}$ on the basis $|k_{12}\rangle$. By similar techniques we can compute the matrix elements of the operator $\hat{L}_{w_{14}}$ on the same basis $|k_{12}\rangle$ and compute the commutator. In the monochromatic case, the computation is particularly simple because the matrix elements of two operators coincide up to a unitary transformation given by the matrix $W_h^{\ph{h}k}=\langle k_{14}|h_{12}\rangle $ (which is a $\{6j\}$ symbol up to a normalization). Therefore we end up computing matrix commutators of $(L_{w_{12}})_h^{\ph{h}h'}$ with $W_{h}^{\ph{h}k} (L_{w_{12}})_k^{\ph{k}k'} W_{k'}^{\ph{k}h'}$. This is non-vanishing already for $j_0=\frac{1}{2}$. Therefore the operators $\hat{L}(\gamma_{w_{12}})$ and $\hat{L}(\gamma_{w_{14}})$ which measure the length of two curves dual to the wedges $w_{12}$ and $w_{14}$ do not commute. In formulae
\begin{equation}
[\,\widehat{L}(\gamma_{\{n,e,e'\}})\,,\,\widehat{L}(\gamma_{\{n,e,e''\}})\,]\neq 0\qquad \textrm{for}\;\;e'\neq e''.
	\label{eq:[L,L']}
\end{equation} 
As we will discuss in section \ref{sec:extended length} the two curves $\gamma_{w_{12}}$ and $\gamma_{w_{14}}$ have the property of intersecting at a point.

Now let's consider two wedges like $w_{12}$ and $w_{34}$ which share a node but do not share any link. We will refer to them as \emph{opposite} wedges. In the four-valent monochromatic case the operators $\hat{L}_{w_{12}}$ and $\hat{L}_{w_{34}}$ defined on $\mc{V}_0^{(j_0 j_0 j_0 j_0 )}$ trivially commute as their matrix elements on the basis $|k_{12}\rangle$ coincide. However this property does not extend to the operators on $\mc{K}_0(\Gamma)$. This can be checked explicitly looking at a non-monochromatic sector. The lowest-dimensional non-trivial case appears to be the non-monochromatic case with dimension $K\geq 4$, as for instance for $\mc{V}_0^{(\frac{3}{2}\, 2\, \frac{3}{2}\, 2 )}$. As a result the operators $\hat{L}(\gamma_{w_{12}})$ and $\hat{L}(\gamma_{w_{34}})$ which measure the length of two non-touching curves defined on the boundary of the same region $R_n$ do not commute
\begin{equation}
[\,\widehat{L}(\gamma_{\{n,e,e'\}})\,,\,\widehat{L}(\gamma_{\{n,e'',e'''\}})\,]\neq 0\qquad \textrm{for}\;\;\{e,e'\}\neq\{e'',e'''\}\;\;.
	\label{eq:[L,L'']}
\end{equation}  
The fact that operators measuring the geometry of space turn out not to commute may appear surprising. For a throughout discussion of the classical origin of this fact we refer to \cite{Ashtekar:1998ak}.

\subsection{Semiclassical behaviour}\label{sec:semiclassical}
Identifying semiclassical states in Loop Quantum Gravity is a difficult issue (see  \cite{Ashtekar:1992tm},\cite{Sahlmann:2001nv},\cite{Giesel:2006uj,Giesel:2006uk,Giesel:2006um}, \cite{Rovelli:2005yj,Bianchi:2006uf,Livine:2006it,Alesci:2007tx,Alesci:2007tg,Bianchi:2007vf} for different attempts). As a tentative attitude, we can restrict attention to the kinematical level and look for states peaked on a given classical $3$-geometry. Here we will adopt a more modest position. We focus on the Hilbert space $\mc{K}_0(\Gamma)$ spanned by spin~network states with graph $\Gamma$ and look for states $\Psi_{\Gamma,c}$ which have the following two  properties: (a) they are required to be peaked on a given expectation value of the geometric operators available on $\mc{K}_0(\Gamma)$ and (b) have vanishing relative uncertainty. Due to the presence in Loop Quantum Gravity of non-commuting geometric operators, these two requirements are already highly non-trivial. 

Let's take into account only the area operator and the angle operator. The angle operator was introduced by Major in \cite{Major:1999mc}. It beautifully fits in the picture of section \ref{sec:picture}. In the language of this paper, it measures the angle between the normals to two intersecting surfaces $S_{e_1}$ and $S_{e_2}$ dual to the links $e_1$ and $e_2$ sharing the node $n$. As a result we have an angle operator for each wedge of $\Gamma$. The angle operator probes the geometry of the spin~network state in the same way as -- at the classical level -- the angle between the normals to two intersecting surfaces depends on the metric of the $3$-manifold. It has the following properties: (i) it commutes with the area operator $\hat{A}(S_e)$ for surfaces dual to links\footnote{In \cite{Ashtekar:1998ak} a non-commutative structure for area operators was shown: area operators for surfaces which \emph{intersect} a node of $\Gamma$ do not commute. Such operators do not appear to fit into the dual picture associated to the graph of a spin~network state. On the other hand, they turn out to be strictly related to the angle operators of \cite{Major:1999mc}.}; (ii) for a four-valent node, the angle operator associated to the wedge $w_{12}=\{n,e_1,e_2\}$ is diagonalized by a spin~network state with the node $n$ labeled by the intertwiner basis $|k_{12}\rangle$ of (\ref{eq:k12}), (iii) the angle operators for wedges sharing a node and a link as $w_{12}=\{n,e_1,e_2\}$ and $w_{14}=\{n,e_1,e_4\}$ do not commute. 

\begin{figure}[t]
\centering
\begin{minipage}[b]{0.4\textwidth}
\begin{tabular}{c}
\includegraphics[height=.6\textwidth]{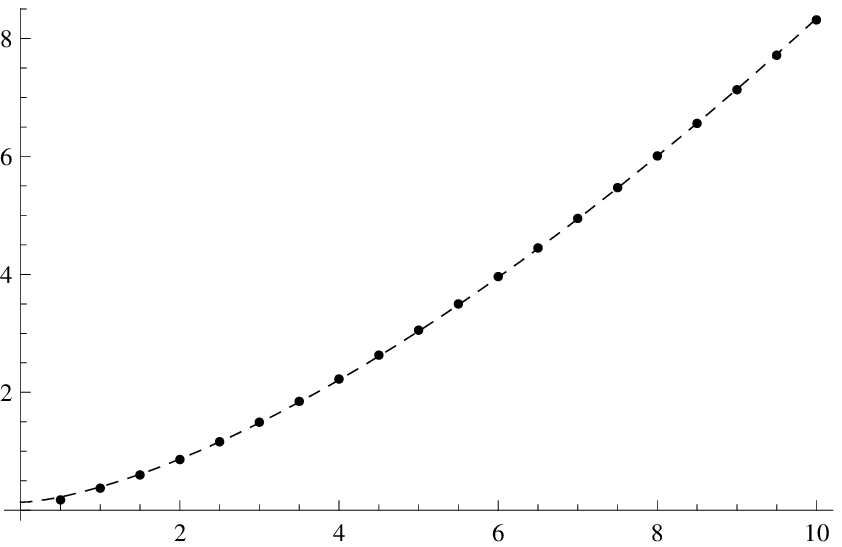}\\
(a)
\end{tabular}
\end{minipage}
\begin{minipage}[b]{0.1\textwidth}
\hspace{1em}
\end{minipage}
\begin{minipage}[b]{0.4\textwidth}
\begin{tabular}{c}
\includegraphics[height=.6\textwidth]{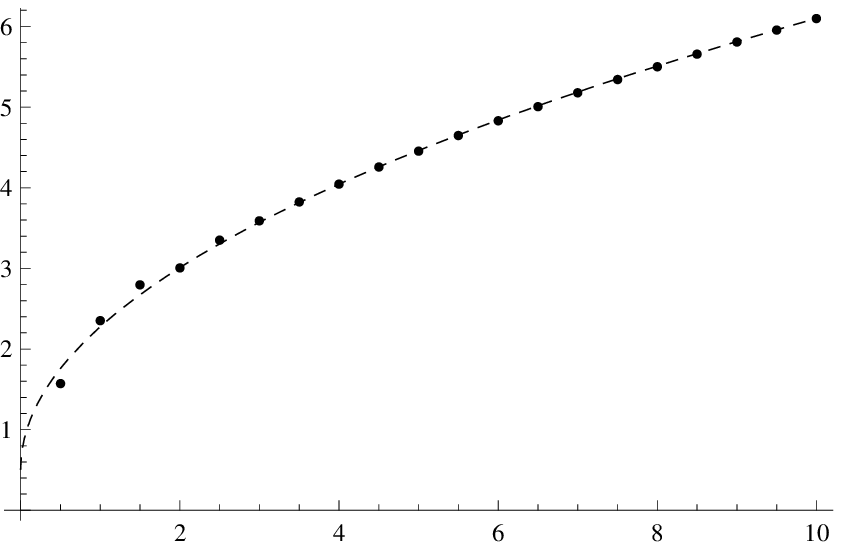}\\
(b)
\end{tabular}
\end{minipage}
\label{fig:volume and length average}
\caption{(a) Expectation value of the volume operator on the semiclassical state (\ref{eq:semiclassical tetrahedron}). Units $(8 \pi \gamma L_P^2)^{3/2}$. Plot as a function of $j_0$. The dashed line is a fit with a curve scaling as $j_0^{3/2}$. (b) Expectation value of the length operator on the semiclassical state (\ref{eq:semiclassical tetrahedron}). Units $(8 \pi \gamma L_P^2)^{1/2}$. Plot as a function of $j_0$. The dashed line is a fit with a curve scaling as $\sqrt{j_0}$.}
\end{figure}

A state $\Psi_{\Gamma,c}\in \mc{K}_0(\Gamma)$ satisfying the semiclassical requirements (a) and (b) can be found taking large spins on the links of $\Gamma$ and labeling four-valent nodes with the \emph{semiclassical intertwiner}
\begin{equation}
	|c\rangle={\sum}_k\; c_k(j_1\mdots j_4) \;|k_{12}\rangle
	\label{eq:|c>}
\end{equation}
introduced by Rovelli and Speziale in \cite{Rovelli:2006fw}. The coefficients $c_k$ are chosen so that the angle operators for the wedges $w_{12}$ and $w_{14}$ are both peaked -- with vanishing relative uncertainties -- around assigned values of dihedral angles $\theta_{12}$ and $\theta_{14}$ corresponding to the geometry of a classical flat tetrahedron. In general, the coefficients $c_k$ have the form of a Gaussian times a phase in $k$. The Gaussian is peaked on a value $k_0$ and has width of order $\sqrt{k_0}$. The  case of a semiclassical regular tetrahedron corresponds to taking $j_1=j_2=j_3=j_4\equiv j_0\gg 1$ (i.e. an eigenstate of the area of the four surface $S_{e_1}\mdots S_{e_1}$ corresponding to large equal eigenvalue) and the coefficients $c_k$ given by
\begin{equation}
	c_k(j_0)=\frac{\sqrt{3}^{1/4}}{(\pi k_0)^{1/4}}\; e^{-\frac{\sqrt{3}}{2} \frac{(k-k_0)^2}{k_0}}\;e^{i \phi_0 k}\;.
	\label{eq:semiclassical tetrahedron}
\end{equation}
The mean value of $k$ is given as a function of $j_0$ by $k_0=\frac{2}{\sqrt{3}}j_0$; the phase is $\phi_0=\frac{\pi}{2}$.
As a test of goodness of this semiclassical state, in figure \ref{fig:volume and length average}-(a) we report the expectation value of the elementary volume operator as a function of $j_0$. It scales as $j_0^{3/2}$. Recalling that the eigenvalues of the area are given by $8\pi \gamma L_P^2\,\sqrt{j_0(j_0+1)}$, we have that the volume scales as the power $\frac{3}{2}$ of the area as expected for a regular classical tetrahedron. Therefore the state obtained superposing spin~network states with coefficients (\ref{eq:semiclassical tetrahedron}) actually appears to describe a configuration such that the chunk of space dual to the node $n$ is peaked on the classical geometry of a flat regular tetrahedron\footnote{Remarkably, beyond this semiclassical situation, the chunk of space associated to a four-valent node can be geometrically interpreted as a `quantum tetrahedron' as pointed out by Barbieri in \cite{Barbieri:1997ks}. See also \cite{Baez:1999tk}}. 

The state $\Psi_{\Gamma,c}\in \mc{K}_0(\Gamma)$ discussed above provides a precious setting for testing the semiclassical behaviour of the elementary length operator introduced in this paper. Numerical investigations indicate that the expectation value of the elementary length operator for a wedge of $w$ on the state $\Psi_{\Gamma,c}$ scales as the square root of $j_0$
\begin{equation}
	(\Psi_{\Gamma,c},\widehat{L}(\gamma_{w_{12}})\Psi_{\Gamma,c})=\sum_{k\, h}\; c_k(j_0)^*\;(L_{w_{12}})_k^{\ph{k}h}\;c_h(j_0)\;\sim\sqrt{j_0}\;.
\end{equation}
as reported in figure \ref{fig:volume and length average}-(b). Therefore it scales exactly as the classical length of an edge of the tetrahedron, i.e. as the square root of the area of one of its faces. Moreover, analysing the relative dispersion we have found that it vanishes as $j_0^{-1/2}$ for large $j_0$. 
 
This same semiclassical analysis can be extended to the case of a non-regular tetrahedron. In that case it would be interesting to show that the operator $\widehat{L}(\gamma_{w_{12}})$ measures exactly the length of the edge $\gamma_{12}$ shared by the triangular faces $S_1$ and $S_2$ of the tetrahedron. A preliminary analysis has been performed and it confirms this expectation. Moreover it sheds a new light on the role played by the inverse-volume operator (section \ref{sec:inverse volume}) in the definition of the length operator. As a test case we have considered a semiclassical state of the form (\ref{eq:|c>}) such that it is peaked on the classical geometry  of a flat tetrahedron with the length of five edges equal to $L_0$ and the length of the remaining edge $\gamma_{34}$ equal to $L_{34}=\alpha L_0$, with $0<\alpha\lesssim 1$. For smaller and smaller $\alpha$ we find that the expectation value of the volume operator goes to zero as expected classically\footnote{A preliminary analysis shows that for smaller and smaller $\alpha$ the dispersion $\Delta V$ grows. This indicates that it is meaningless to simply take $\alpha\to 0$ as the state is not any more a good semiclassical state and we are entering in a deep quantum regime.}. On the other hand the expectation value of the length operator for the edge $\gamma_{12}$ (i.e. the edge opposite to $\gamma_{34}$) does not grow, remains \emph{finite} and equal to the value $L_0$.

This set of results strongly strengthens the relation between the quantum geometry of a spin~network state  and the classical simplicial geometry of piecewise-flat $3$-metrics pointed out by Immirzi, by Barbieri, by Baez and by Barrett \cite{Immirzi:1996dr,Barbieri:1997ks,Baez:1999tk,Baez:1997zt}. Notice that this relation plays a key role in the analysis of the graviton propagator in Loop Quantum Gravity \cite{Rovelli:2005yj,Bianchi:2006uf,Livine:2006it,Alesci:2007tx,Alesci:2007tg,Bianchi:2007vf}.

Let's also notice that testing Thiemann's proposal for the length operator \cite{Thiemann:1996at} on the semiclassical state discussed in this section would hopefully clarify the relation between the two proposals.

\section{The length of an extended curve}\label{sec:extended length}
In the previous section we have investigated the properties of the `elementary' length operator defined on $\mc{K}_0(\Gamma)$ for a curve dual to a wedge of the graph $\Gamma$. In this section we discuss two natural 
extensions which lead to a tentative definition of the operator measuring the length of an extended curve in $\Sigma$. These two extensions bring with them interesting stronger requirements for semiclassical states.\\

Let's start setting the framework. We work on the Hilbert space $\mc{K}_0(\Gamma)$ spanned by spin~networks with graph $\Gamma$. We assume that the graph $\Gamma$ is dual to a cellular decomposition\footnote{A cellular decomposition is a decomposition of a manifold in terms of open balls. In general, the dual of a graph $\Gamma$ is a cellular complex $\Gamma^*$ which however does not provide a decomposition in cells of the $3$-manifold $\Sigma$.} 
$\mc{C}_\Sigma$ of the $3$-manifold $\Sigma$ \cite{RouSan:2005rh,Oeckl:2005rh}. We call $\mc{C}_\Sigma(\Gamma)$ the cellular decomposition dual to the graph $\Gamma$, $\mc{C}_\Sigma(\Gamma)\sim \Gamma^*$. It contains $3$-cells $R_n$ dual to the nodes of $\Gamma$ and $2$-cells $S_e$ dual to the links of $\Gamma$. Moreover it contains $1$-cells $\gamma$ which we call elementary curves and $0$-cells $v$ which we call vertices. We call \emph{dual network} the set of elementary curves of $\mc{C}_\Sigma$.

We also need the notion of a $2$-complex $\bar{\Gamma}$ associated to the graph $\Gamma$. It is defined so that it contains $\Gamma$ and is made by the set of surfaces $f$ dual to $1$-cells $\gamma$ of $\mc{C}_\Sigma(\Gamma)$. The boundary of a surfaces $f\subset\bar{\Gamma}$ is given by a collection of nodes and links of $\Gamma$. We call it a \emph{plaquette} and indicate it with $p$. 
The $2$-complex $\bar{\Gamma}$ can be easily visualized in the following way. Imagine we dip and remove a wire model of the spin~network graph into a soap solution. What we get is a collection of soapfilm surfaces $f$ which has the `wire' $\Gamma$ as boundary. Each soapfilm surface has as boundary a plaquette of the wire. 

Notice that the $2$-complex $\bar{\Gamma}$ describes a cellular decomposition of the $3$-manifold $\Sigma$. Such decomposition has nodes of $\Gamma$ as $0$-cells and links of $\Gamma$ as $1$-cells. Moreover it has $2$-cells given by the soapfilm surfaces $f$ and  $3$-cells which we call \emph{bubbles} and indicate with $b$. A bubble of the $2$-complex $\bar{\Gamma}$ is dual to a vertex $v$ of the cellular decomposition  $\mc{C}_\Sigma(\Gamma)$. In the following we make use of this nomenclature which is summarized in table \ref{tab:dual}.

\begin{table}
\centering
\begin{tabular}{ccc}
& $\bar{\Gamma}$  & $\mc{C}_\Sigma(\Gamma)$\\
\hline
$0$-cells  & nodes $n$  &  vertices $v$\\
$1$-cells  & links $e$  &  curves $\gamma$\\
$2$-cells  & films $f$  &  surfaces $S$\\
$3$-cells  & bubbles $b$  &  regions $R$
\end{tabular}
	\label{tab:dual}
	\caption{Cells of the decompositions $\bar{\Gamma}$ and $\mc{C}_\Sigma(\Gamma)$. The $k$-cells of $\bar{\Gamma}$ are dual to $(3-k)$-cells $\mc{C}_\Sigma(\Gamma)$. The graph $\Gamma$ is given by the collection of $0$-cells and $1$-cells of $\bar{\Gamma}$. In the text, the collection of nodes and links of $\Gamma$ which are given by the boundary of a film $f$ is called a plaquette, $p=\p f$. We have improperly used the same name bubble both for a $3$-cell and for its boundary given by a collection of films $f$.}
\end{table}

\subsection{The length of a curve dual to a plaquette: a continuous-metric condition}\label{sec:continuous-metric}
In section \ref{sec:properties} we have discussed the properties of the elementary length operator associated to a wedge of the graph $\Gamma$. Notice that a plaquette of $\Gamma$ is given by a collection of wedges $w_\alpha$ so that $p=\{w_1\mdots w_P\}$. Moreover, being the boundary of a surface $f$ of the $2$-complex $\bar{\Gamma}$, a plaquette is naturally associated to an elementary curve dual to $f$. We call such curve either $\gamma_f$ or equivalently $\gamma_p$. In this section we investigate the possibility of extending the elementary length operator for a wedge to a length operator for an elementary curve dual to a plaquette, $\widehat{L}(\gamma_p)$. \\

Thanks to the results of section \ref{sec:commutators}, we know that length operators associated to different wedges belonging to the same plaquette commute. Therefore they can be simultaneously diagonalized. Moreover, as they commute with the area operator, we can choose a specific configuration for the spin labels. 

Let's start considering the particular situation when all the links of $\Gamma$ ending at a node of the plaquette are labelled by spins which are all equal to $j_0$. A simultaneous eigenstate of all the operators $\widehat{L}(\gamma_{w_1})\mdots \widehat{L}(\gamma_{w_P})$ can be easily found. We simply have to choose at each node of the plaquette one of the $2j_0+1$ intertwiners $|(l_i)_{w_\alpha}\rangle$ which diagonalizes the $(2j_0+1)\times(2j_0+1)$ matrix $(L_{w_\alpha})_k^{\ph{k}h}$. In this way we have defined a spin~network state which is a simultaneous eigenstate of each of the operators $\widehat{L}(\gamma_{w_1})\mdots \widehat{L}(\gamma_{w_P})$ to eigenvalues $(8\pi \gamma L_P^2)^\frac{1}{2} \,l_i$ which are in general different. From the point of view of the cellular decomposition $\mc{C}_\Sigma(\Gamma)$ this means that this state describes a quantum geometry such that the curve dual to the plaquette $p$ has a different length when seen from different wedges. 
This fact has a flavor which is very close to the simplicial discontinuous-metrics discussed in \cite{Barrett:2000jy,Wainwright:2004yn,Dittrich:2008va}. 

The possibility of imposing a \emph{continuous-metrics condition} can be considered. For instance if in the spin~network state described above we choose the same intertwiner $|(l_{\bar{i}})_{w_\alpha}\rangle$ in each of the nodes of the plaquette, then we have that the curve $\gamma_p$ has the same length $(8\pi \gamma L_P^2)^\frac{1}{2} \,l_{\bar{i}} $ when seen from different wedges. Let's call this state $\Psi_{\{p,j_0,l_{\bar{i}}\}}$. In this case the continuous-metrics condition would involve all the nodes belonging to the plaquette, not simply single nodes. In this sense it can be called a \emph{local} condition, but not an ultralocal one.

Notice that for general fixed spins $j_e$ labelling the links of $\Gamma$, the spectra of the matrices $(L_{w_1})_k^{\ph{k}h}$,.., $(L_{w_P})_k^{\ph{k}h}$ do not match. This is due to the fact that the spectrum for the wedge $w=\{n,e,e'\}$ depends on the spins labelling the links which end at the node $n$ and different wedges of the plaquette may have different spins. In this case the requirement of a matching of eigenvalues is highly non-trivial and in general cannot be satisfied. However, this does not mean that an operator $\widehat{L}(\gamma_p)$ properly associated to the length of the curve dual to the plaquette $p$ cannot be introduced. In fact the operator has to be defined on the Hilbert space $\mc{K}_0(\Gamma)$ and not on a subspace with prescribed spins on the links. This suggests that an operator $\widehat{L}(\gamma_p)$ satisfying the continuous-metrics condition can be introduced, that it acts locally on the plaquette both on spins and intertwiners and that the $2j_0+1$ states $\Psi_{\{p,j_0,l_{\bar{i}}\}}$ described above are eigenstates of $\widehat{L}(\gamma_p)$. 

A second milder possibility is to impose the continuous-metrics condition only on semiclassical states. With respect to the discussion of section \ref{sec:semiclassical}, this amounts to add to the semiclassicality requirements (a) and (b) the third requirement (c) that the expectation value of length operators for different wedges belonging to the same plaquette match. Notice that in order to satisfy (a) and (b) an `ultralocal' superposition of spin~network states is enough. That is we have a product of coefficients $c_k$ as in (\ref{eq:|c>}), one for each node and not relatively constrained. In order to satisfy (c) too we need a superposition which is \emph{local} with respect to the plaquette but in general not ultralocal. The superposition over intertwiners labeling nodes of the plaquette may be required to be of a non-product (or entangled) form or more simply to be a product of non-independent coefficients. These issues clearly deserve further work.

\subsection{The length of a curve belonging to the dual network}
Finally we describe how to define the length operator for an `extended' curve $\gamma$ embedded in the $3$-manifold $\Sigma$ out of the `elementary' length operator discussed in section \ref{sec:properties}. We make use of the framework set at the beginning of this section and ultimately of the dual picture of quantum geometry described at the very beginning of this paper in section \ref{sec:picture}.\\ 

The idea is to introduce a length operator which measures the length of a curve belonging to the dual~network. We recall that starting from the spin~network graph $\Gamma$ we can identify a cellular decomposition $\mc{C}_\Sigma(\Gamma)$ of the $3$-manifold $\Sigma$ which is dual to the graph $\Gamma$. A curve belonging to the dual~network is simply defined as a path on the $1$-skeleton of $\mc{C}_\Sigma(\Gamma)$. In the language of the soapfilm-like $2$-complex $\bar{\Gamma}$, it is a curve which goes from bubble to bubble crossing common faces of the bubbles. Such curve is the composition of elementary curves, $\gamma=\gamma_{f_1}\!\circ \cdot\cdot\circ\, \gamma_{f_N}$. Therefore we can attempt to use the `elementary' length operator to define the length of the extended curve as a sum of elementary lengths in a fashion analogous to what is done for the volume operator

Given the discussion of the previous subsection, we have to declare if we are attempting to impose a continuous-metrics condition or not. Even if the possibility of imposing such condition can be considered, in the following we will avoid relying on it. Therefore, when discussing the length of the curve $\gamma$, we have to tell from which wedge the portion $\gamma_f$ is seen. We will indicate it explicitly writing $\gamma$ in terms of $\gamma_w$ for wedges. Within the dual picture given by the cellular complex $\mc{C}_\Sigma(\Gamma)$, what we have to give is the curve $\gamma$ together with a tube of $3$-cells which contains $\gamma$ on its boundary. We call such tube $t_\gamma$.

Thus let's consider the curve $\gamma=\gamma_{w_1}\!\circ \cdot\cdot\circ\, \gamma_{w_N}$. We have that, in order to define the length operator $\widehat{L}(\gamma)$ as a sum of $\widehat{L}(\gamma_{w_i})$, the elementary length operators $\widehat{L}(\gamma_{w_1}), \ldots,\widehat{L}(\gamma_{w_N})$ have to be compatible observables. As the length operators for two wedges $w_1$, $w_2$ sharing a node do not commute (see section \ref{sec:commutators}), we have that only a specific class of nice curves can be considered. Within the dual picture we have that the curve $\gamma$ cannot contain two elementary curves $\gamma_{w_1}$ and $\gamma_{w_2}$ which belong to the same $3$-cell $R$ of $\mc{C}_\Sigma(\Gamma)$. Therefore the curve $\gamma$ cannot be `too spiky' and cannot be self-intersecting or `almost' self-intersecting. This latter condition comes from the fact that two elementary curves $\gamma_{w_1}$ and $\gamma_{w_2}$ can belong to the same $3$-cell $R$ and nevertheless do not intersect. As an example, for a four-valent node we have a $3$-cell $R$ with the topology of a tetrahedron and two opposite edges of the tetrahedron do not intersect.

For the class of nice curves described above, the length operator defined on $\mc{K}_0(\Gamma)$ is simply given by the following sum
\begin{equation}
	\widehat{L}(\gamma_{w_1}\!\circ \cdot\cdot\circ\, \gamma_{w_N})=\widehat{L}(\gamma_{w_1})+\cdots+\widehat{L}(\gamma_{w_N})\;.
	\label{eq:extended length}
\end{equation}
At this point we can consider two nice curves $\gamma$ and $\gamma'$ and consider the two operators measuring their lengths. We have that they do not commute if the two curves intersect or almost-intersect. This result can be restated in terms of tubes $t_\gamma$, $t_{\gamma'}$ of the cellular decomposition $\mc{C}_\Sigma(\Gamma)$ as
\begin{equation}
	[\;\widehat{L}(\gamma)\,,\,\widehat{L}(\gamma')\;]\neq 0 \qquad\textrm{for}\quad t_\gamma\; \cap\; t_{\gamma'} \neq \emptyset\;\;.
  \label{eq:intersecting curves}
\end{equation}
Otherwise, if $t_\gamma\; \cap\; t_{\gamma'} = \emptyset$ then the two operators commute. 

As a consequence of the non-commutativity of length operators, we have that we cannot `prepare' a state which describes a dual~network with definite lengths. Finding a semiclassical state for the dual~network along the lines discussed in sections \ref{sec:semiclassical} and \ref{sec:continuous-metric} would be extremely interesting. Work is in progress in this direction.

\section{Conclusions}\label{sec:conclusion}
An elementary attempt to quantize general relativity in the connection representation would lead to an electric field operator acting as a functional derivative
\begin{equation}
	\parbox{145pt}{``$\;\;\displaystyle \widehat{E}^a_i(x)=-i\, 8\pi \gamma L_P^2\, \frac{\de}{\de A_a^i(x)}\;\;$''}.
\end{equation}
Within such framework, promoting a classical geometrical quantity like the volume of a region (see equation \ref{eq:vol}) to a quantum operator appears out of hope. From this perspective, the fact that -- in the loop approach -- geometric operators corresponding to areas and volumes can be introduced and studied is a highly non-trivial result. In this paper we have introduced in Loop Quantum Gravity a new geometric operator -- the length operator. As for the area and for the volume, the length operator turns out to have a discrete spectrum. We have avoided entering into the difficult issue of observability of such spectrum. We refer to \cite{Rovelli:2004tv,Thiemann:book2007,Ashtekar:2004eh} for perspectives and to \cite{Dittrich:2007th,Rovelli:2007ep} for a recent discussion. 

The operator is constructed starting from the classical expression (\ref{eq:classical L})--(\ref{eq:classical G}) for the length of a curve. The quantization procedure goes through an \emph{external} regularization of the classical quantity, a canonical quantization of the regularized expression and an analysis of the existence of the limit in the Hilbert space topology. The operator constructed in this way has a number of properties which are discussed in sections \ref{sec:properties} and \ref{sec:extended length}. Let's highlight some of them:
\begin{itemize}
	\item The length operator fits into the dual picture of quantum geometry proper of Loop Quantum Gravity. More in detail, for given spin~network graph, the operator measures the length of a curve in the dual~graph.
	\item An `elementary' length operator can be introduced. It measures the length of a curve defined in the following intrinsic way. A node of the spin~network graph is dual to a region of space and a couple of links at such node are dual to two surfaces which intersect at a curve. The `elementary' length operator measures the length of this curve.
	\item The `elementary' length operator has a discrete spectrum. A number of eigenstates and eigenvalues computed algebraically and numerically are presented in section \ref{sec:spectrum}. 
	\item The `elementary' length operator has non-trivial commutators with other geometric operators, as shown in section \ref{sec:commutators}.
	\item A semiclassical analysis has been performed. It shows that the `elementary' length operator has the appropriate semiclassical behaviour: on a state peaked on the geometry of a classical tetrahedron it measures the length of one of its edges. 
  \item	For given spin~network graph, the length operator for a curve in the dual~graph can be written in terms of a sum of `elementary' length operators. 
  \item The analysis of the length of a curve in the dual graph shows that the geometry described by a spin~network state is in general discontinuous. The possibility of identifying states which describe a continuous geometry is discussed in section \ref{sec:continuous-metric}.
  \item The length operators for two curves in the dual~network do not commute if the two curves come close to intersecting.
\end{itemize}
There remains still much to be done. Mainly, an analysis of the extension of the operator to the full Hilbert space $\mc{K}_0$, a study of the continuous-metrics condition of section \ref{sec:continuous-metric} and an investigation of its consequences on the semiclassical behaviour of Loop Quantum Gravity.


\section*{Acknowledgments}
\hspace{1.5em}Thanks to Carlo Rovelli for enlightening discussions. I wish also to thank Daniele Oriti, Leonardo Modesto, Simone Speziale and Alejandro Perez for helpful comments and suggestions. Support by Della~Riccia Foundation is gratefully acknowledged.



\providecommand{\href}[2]{#2}\begingroup\raggedright\endgroup


\end{document}